\documentclass[12pt]{article}
\usepackage[dvips]{graphicx}
\usepackage{amsmath,amssymb}

\begin{document}
\title{ Long-range  correlation of neutrino in  pion
decay   } 
\author{ K. Ishikawa and  Y. Tobita}
\maketitle
\begin{center}
  Department of Physics, Faculty of Science, \\
Hokkaido University Sapporo 060-0810, Japan
\end{center}


\begin{abstract}
 Position-dependent property of the neutrino produced in high-energy pion 
decay is
 studied theoretically using  a wave packet formalism. It is found that the
 neutrino probability  has a universal finite-distance  correction if
 the pion has a long coherence length. 
This 
correction has origins in the interference of the neutrino that is
 caused by the small  neutrino mass and a light-cone 
singularity of the pion and muon system. 
This term is positive definite and leads excess to the neutrino flux 
over the in-coherent value at macroscopic distance of  near detector 
region. The flux decreases slowly with distance in a universal
 manner that is 
determined by the absolute value of the neutrino mass and energy. 
Absolute value of the neutrino mass is expected to be measured
 indirectly from  the neutrino detection probability at finite distance.

\end{abstract}

\newpage

\section{Introduction}

Coherence phenomenon of the pion decay process becomes important when
the pion is a wave of the long coherence length. The decay
process then is studied  using a time-dependent Schr\"{o}dinger
equation, and
the neutrino in the final state becomes  a wave that
maintains the coherence  for a long interval. Furthermore if this
neutrino wave is a 
superposition of waves of particular complex phases,  this shows an 
interference phenomenon. The total 
number of the neutrino events is  large  in the recent ground
experiments, and the interference of the neutrino may become observable there.
To  study  the interference 
phenomenon of the neutrino in the decay of high-energy pion 
\begin{eqnarray}
\pi \rightarrow \mu+\nu_{\mu},
\label{pion-decay}
\end{eqnarray}
especially the physics derived from the 
time-dependent wave function and  sensitive to absolute
value of the neutrino mass, is the subject of the present paper.

Understanding the neutrino is in rapid progress. Neutrinos are very light 
and the  mass squared differences were given 
from recent flavour oscillation
experiments~\cite{SK-Atom,SK-Solar,SNO-NC,KamLAND-Reactor,Borexino,K2K}
 using neutrinos from the sun, accelerators, reactors, and atmosphere as,   
\cite{particle-data}
\begin{align}
\Delta m^2_{21} &= m_2^2 - m_1^2 = (7.59 \pm 0.20)\times 10^{-5}~[\text{eV}^2/c^4],\nonumber\\
|\Delta m^2_{32}| &= |m_3^2 - m_2^2|= (2.43 \pm 0.13)\times 10^{-3}~[\text{eV}^2 /c^4],
\end{align}
where $m_i~(i=1-3)$ are mass values.
The absolute
values of masses are important but are not found from the oscillation
experiments. Tritium beta decays \cite{Tritium}~have
been used for determining the 
absolute value but the existing upper bound for the effective electron 
neutrino squared-mass  is of order  $  2 ~[\text{eV}^2/c^4]$ and the mass 
is $ 0.2 - 2 ~ [\text{eV}/c^2]$ from cosmological
observations~\cite{WMAP-neutrino}. 
The neutrino is also idealistic
 for the test of  quantum mechanics due to its  weak interaction with matters, 
although the validity of quantum mechanics is certain from many tests
and applications made in the past. 
For  these implications, much works should be done
further.

To study the coherence phenomenon of the neutrino, coherence properties 
of the pion should be known. One particle states which have finite
coherence lengths are described by wave packets 
\cite{Ishikawa-Shimomura,Ishikawa-Tobita-ptp,Ishikawa-Tobita}.
By using wave packets, position-dependent amplitudes and probabilities  
are also computed.
{\samepage\footnote{The general arguments about
the wave packet scattering are given in
\cite{Goldberger,newton,taylor,Sasakawa}. In these works, however, situations
where wave packet effects are negligible were  considered mainly.}}
Coherence properties of the neutrino produced in pion 
decay are determined by the pion's coherence properties and decay
dynamics. When the pion's coherence length is short, the pion and the neutrino
 behave like particles. The total decay rate and average life time are
 computed in the usual way and the neutrino  
shows no physical phenomenon which has an origin in the spatial coherence. 
 In other situation when the pion has a long coherence length, the pion and the
 neutrino behave like waves and the interference phenomenon of neutrino 
appears in the space-dependent wave function and probability, which  will 
be presented here.

We focus on  quantitative analysis  of the neutrino  
produced in  pion decay. If these  particles 
are described by plane waves, the transition amplitude has the delta
function of the energy and momentum conservation and the total
transition probability is proportional to time interval T.
The decay rate and the mean  life time are found from the  
proportional constant. Here we study finite-size corrections in the pion
decay process. Namely the various probabilities of the pion decay
process are computed in the situation where the distance between the
initial and finial states is finite and the time interval between
them is also finite. 
From the integral representation of the probability, as is
explicitly given in section 4,  a large  finite-size correction is
lead  if the correlation function of the pion and muon 
system has a light-cone singularity.  
The dependence of the probabilities on these
values are also determined by the light-cone singularity. 
The
time-dependent wave function of the whole process gives rise to this 
correction,   which, furthermore, becomes observable quantity by 
the use of wave packet expression of the particle states, instead 
of the plane waves.

\begin{figure}[t]
 \includegraphics[scale=0.3]{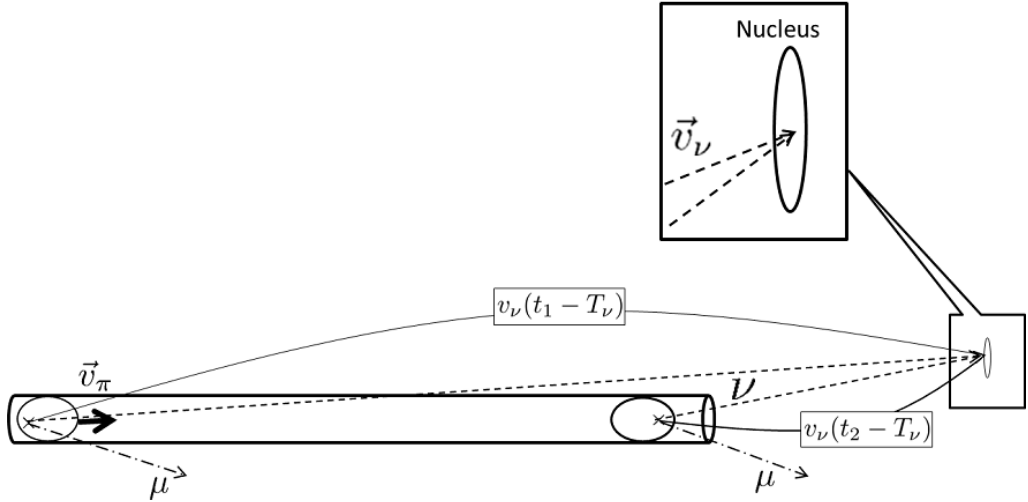}
\caption{The geometry of the neutrino interference experiment. The
 neutrino is observed by the detector at $T_{\nu}$  and produced 
at $t_1$ or $t_2$ in the inside of the traveling pion wave. }
\label{fig:geo}
\end{figure}

A neutrino produced in pion decay propagates finite
distance before it is detected.  The distance is not fixed but varies within 
  the size of the pion decay region. So the neutrino wave at the detector is a
  superposition of those
that are produced at different positions. If these positions are in one
pion, as in Fig.\,$(\ref{fig:geo})$ the interference phenomenon of the 
neutrino wave is possible. Let the
pion's   coherence length,  $\sqrt{\sigma_{\pi}}$, and  velocity,
$v_{\pi}$, and find the criteria when this happens. When  the neutrino
is produced either at  time $t_1$ or $t_2$ from the pion that is prepared 
at $T_{\pi}$ and is detected 
at $T_{\nu}$, after traveling with the light velocity,  
they satisfy
\begin{eqnarray}
|(c(T_{\nu}-t_1)+v_{\pi}(t_1-T_{\pi}))-(c(T_{\nu}-t_2)+v_{\pi}(t_2-T_{\pi}))|
 \leq \sqrt{ \sigma_{\pi}},
\label{coherence-condition}
\end{eqnarray}
if these particles travel in one-dimensional space, where the light
velocity is used for the neutrino velocity $v_{\nu}=c$.
So this is one of the conditions for the neutrino interference in the
one-dimensional space. 
If this criteria holds, the neutrino wave from $t_1$ can interfere with 
the neutrino wave from $t_2$. At high energy, the pion velocity is
close to the light velocity and the left hand side of
Eq.\,$(\ref{coherence-condition})$ becomes to
$c(m_{\pi}^2/2E_{\pi}^2)(t_1-t_2)$, 
hence this condition is satisfied at the 
time difference $c(t_1-t_2) \approx
\sqrt{\sigma_{\pi}}(2E_{\pi}^2/m_{\pi}^2)$.
 When this length $c(t_1-t_2)$ is macroscopic size, the
coherence phenomenon of the neutrino at the macroscopic length is possible.
We will estimate the coherence 
lengths of these particles 
in second
section and will see that this condition in three-dimensional space 
is satisfied in the macroscopic
distance.  
In this situation the amplitude for detecting neutrino at a finite
distance is given by the integral of the product of wave
functions of the pion, muon, and neutrino over the coordinate.
Now we recall that the velocity of  any relativistic waves approaches 
the light velocity, when the momentum becomes infinite. Hence the  space-time positions
of weak interaction $(t_i,{\vec x}_i);~i=1,2$ where the neutrino is produced 
in a system
of the incoming pion wave and outgoing muon wave have the light velocity at the
large momentum and the two point correlation function defined from the 
pion and muon states in the decay probability $\Delta_{\pi,\mu}(t_1-t_2,{\vec
x}_1-{\vec x}_2)$ has a light-cone singularity.
Since the light-cone singularity  is real and extended in wide area of 
space and time which almost overlaps with the neutrino's space-time
path, the neutrino probability, which is the integral of the neutrino wave
function multiplied by the light-cone singularity, gets the direct effect 
from the phase of the neutrino wave function. 
The phase of  the neutrino wave of mass $m_{\nu}$ and energy $E_{\nu}$
is expressed by using the differences of two positions $\Delta {\vec x}$
and of two times $\Delta t$, as $\exp{(i \phi)}$, where $\phi$ is defined
by $E_{\nu}\Delta t-{\vec p}_\nu\cdot{\Delta \vec x}$. This phase  is expressed
%
%
%
in the form $\phi=(m_{\nu}^2/2 E_{\nu}) \times  c \Delta t $, if
 $ \Delta {\vec x}=c{\vec n}_{\nu}  \Delta t$ is substituted.
Here ${\vec n}_{\nu}$ is the unit vector along the neutrino momentum. 
Since the angular velocity ${m_{\nu}^2/2 E_{\nu}}$ is extremely
small, this gives a large and long-distance correction.  
Thus the light-cone singularity of the pion and muon system and  the light
velocity and slow phase of the neutrino wave function are combined to
produce  the  finite-distance correction to the neutrino
probability. Because 
the neutrino's mass is extremely small, the correction is expected to 
be of the large distance. 

We
investigate the  physical phenomena that have a deep connection  with 
the neutrino's coherence and interferences in  the  high-energy pion
decay. 
For simplicity we
study the parameter region where the absolute value of the neutrino mass is 
much larger than the mass  differences and the  effect  of 
the flavour oscillation is  negligible. More general situations will be
studied later.  
Other interference phenomena caused by solar neutrinos, reactor
neutrinos, and others are studied in a next
work.

This paper is organized in the following manner. In section 2, the sizes
and shapes of wave packets are studied.  In section  3, we
study  neutrino  production amplitude  in the pion decay 
and in section 4, we study neutrinos detection probability. The
probability is expressed as the sum of normal term and new term. The
position-dependent probability is obtained from the light-cone
singularity of the decay correlation function in
section 5. Summary and prospects are given in section 6.

\section{Wave packets }
In the decay of the pion Eq.\,$(\ref{pion-decay})$, the pion is in the
beam and the neutrino is detected by the detector. Pions are produced in
collisions of protons  with nucleus and are superpositions of plane
waves. Weak decay of a pion is studied normally using momentum
eigenstates of the pion, neutrino, and muon. In the present paper,
position dependent probability for the neutrino is studied and the neutrino is
expressed by a wave packet, which is extended   around its centers in 
position  and momentum. In order to clarify this probability,   initial
pion is also expressed by a wave packet.  The wave
packet size is determined by the detection process for the neutrino but
by the production process and medium effect for the pion.       

    
A particle in matter is described by a wave that maintains  coherence in
a finite region defined by the mean free path
\cite{Ishikawa-Tobita-ptp}. When the particle of 
the finite mean free path is emitted into vacuum, the wave representing
this particle propagates. This wave  is  expressed by a wave packet. 
Hence  a quantum mechanical transition of this particle state 
is computed from the overlap integral of the wave functions.  The wave
function of the finite spatial size  has also a finite  momentum 
width. 

We estimate the wave packet size of a proton first and those of a pion
and neutrino next, 
following the method of our previous works  \cite{Ishikawa-Shimomura,Ishikawa-Tobita-ptp}.

\subsection{Size of pion  wave packet }
Since pions are produced in collisions of protons  with nucleus,
the coherence lengths of pions are  determined by those of
a proton and nucleus.  A proton in matter interacts with nucleus and has 
a finite mean free path or a finite coherence length. The
target nucleus has a microscopic size of order $10^{-15}$\,[m] and its
position is extended in a size of nucleus wave function in matter. Its
magnitude is slightly larger than a nucleus intrinsic size. We use in
the present paper the size of the order of $10^{-15}$\,[m] for the nucleus size.

\subsubsection{Proton  mean free path}

The mean free path of a  charged particle 
is determined by its  scattering  with atoms in matter 
by Coulomb interaction. An energy loss is also determined by the same cross
section. Data on the energy loss   are summarized well  in particle data 
summary \cite{particle-data} and are made use of for the evaluation of the   
mean free path of the proton. 

The proton's energy loss rate at the momentum, $1$\,[GeV/c], for several 
metals such as Pb, Fe, and
others are  
\begin{eqnarray}
-{d E \over d x}=1- 2 ~[\text{MeVg}^{-1}\text{cm}^2],   
\end{eqnarray}
hence we have the  mean free path of the $1\,[\text{GeV}/c]$ proton in
the material of the density $\rho$,
\begin{eqnarray}
L_\text{proton}={E \over {dE \over d x}\times \rho }= {1 ~[\text{GeV}]
 \over (1 - 2)\times
 10 ~[\text{MeV g}^{-1} \text{cm}^2 \text{g cm}^{-3}]} = 50 -  100~[\text{cm}].
\label{mfp-proton-1}
\end{eqnarray}
At a lower energy,  $0.2\,[\text{GeV}/c]$, the energy loss rate of the
proton is about
$10\,[\text{MeVg}^{-1}\text{cm}^2]$ and the mean free path is  
\begin{eqnarray}
L_\text{proton}=10\,[\text{cm}].
\end{eqnarray}

The proton maintains coherence within the mean free path. Hence we use
the  mean free path for the coherence length of the proton ${\delta x}_\text{proton}$,
\begin{eqnarray}
{\delta x}_\text{proton}= L_\text{proton}.
\end{eqnarray}

When the proton of the finite coherence length is emitted into the
vacuum or a system composed of dilute gas from the matter,  it has the 
coherence length defined in matter. The proton has a constant coherence 
length in vacuum or in dilute gas when it is moving freely. The
coherence varies when the proton is accelerated. If the potential energy
$\mathcal{V}$ is added to the particle of momentum $p_\text{before}$, then the momentum
becomes $p_\text{after}$ and satisfies 
\begin{eqnarray}
\sqrt{p_\text{before}^2+m^2}+\mathcal{V}=\sqrt{p_\text{after}^2+m^2}.
\label{potential-energy}
\end{eqnarray}
From Eq.\,$(\ref{potential-energy})$ the variants of the momentum satisfy
\begin{eqnarray}
& &v_\text{efore}\times\delta p_\text{before}=v_\text{after}\times\delta p_\text{after},\nonumber\\
& &v_\text{before}=\frac{p_\text{before}}{\sqrt{p_\text{before}^2+m^2}},~
v_\text{after}=\frac{p_\text{after}}{\sqrt{p_\text{after}^2+m^2}}. 
\end{eqnarray}
Hence the coherence length of a particle, ${\delta x}_\text{before}$,
which is proportional to the inverse of ${\delta p}_\text{before}$, becomes
to  ${\delta x}_\text{after}$ after  the acceleration from a velocity 
$v_\text{before }$
to a velocity $v_\text{after}$. The coherence length is determined by the 
velocity ratio,
\begin{eqnarray}
{\delta x}_\text{after}={\delta x}_ \text{before}\times{v_\text{after} \over
 v_\text{before}}.
\end{eqnarray}
The  velocity is bounded
by the light velocity $c$,
and the velocity ratio from $1\,[\text{GeV}/c]$ to $10\,[\text{GeV}/c]$
 is about $1.2$ and that from $0.2\,[\text{GeV}/c]$ to $10\,[\text{GeV}/c]$
 is about $5$. Hence the proton 
 of $10~[\text{GeV}/c]$ regardless of the energy in matter has the mean 
free path
\begin{eqnarray}
{\delta x}_\text{proton}\approx 40 - 100~[\text{cm}],
\label{mfp-proton-2}
\end{eqnarray}   
in vacuum or a dilute gas.

\subsubsection{Pion mean free path }

Coherence length of pions which are produced by a proton collision 
with target nucleus is determined by the proton's  initial 
coherence  length Eq.$(\ref{mfp-proton-2})$ and  the target size $10^{-15}$\,[m], which is negligibly small. Since 
the time interval of the proton wave is the same as that of the pion wave, the coherence length
of the pion, ${\delta
x}_\text{pion}$, is given from that of the proton,
 ${\delta x}_\text{proton}$, in the form
\begin{eqnarray}
\frac{{\delta x}_\text{proton}}{v_\text{proton}} = \frac{{\delta
 x}_\text{pion}}{v_\text{pion}},~{\delta x}_\text{pion}= \frac{v_\text{pion}}{v_\text{proton}}{\delta x}_\text{proton}\approx {\delta x}_\text{proton} .
\end{eqnarray}
In relativistic energy region,
particles have light velocity.
Consequently from Eq.\,$(\ref{mfp-proton-2})$,  we have the pion's coherence
of $1\,[\text{GeV}/c]$ or larger 
momentum 
\begin{eqnarray}
{\delta x}_\text{pion}\approx 40 - 100~[\text{cm}].
\label{mfp-pion}
\end{eqnarray}   
We use this  value of  
Eq.\,$(\ref{mfp-pion})$ as the size of the wave packet 
\begin{eqnarray}
\sqrt{\sigma_{\pi}}={\delta x}_\text{pion},
\end{eqnarray}
in latter sections.  

In vacuum and dilute gas, pions propagate freely with the same coherence 
lengths.   From
Eqs.\,$(\ref{mfp-proton-1})$,\,$(\ref{mfp-proton-2})$,\,$(\ref{mfp-pion})$, the
proton and pion have the coherence lengths of the order $50 - 100$\,[cm].

\subsection{Size of neutrino    wave packet}
A  size of wave packet for observed neutrino  is determined 
from its detection process.  Neutrinos interact with nucleus or 
electrons in atoms incoherently and the neutrino event is identified by
the reaction process in detector. Hence the 
size of the observed neutrino is not determined by the mean free path 
but by a  size of the minimum physical system that the neutrino
interacts and gives the
signal. They are either  nucleus or electrons in atom.  The nucleus have 
sizes of order $10^{-15}$\,[m] and the electron's wave functions have 
sizes of order $10^{-10}$\,[m].  So neutrino wave packet is either $10^{-10}$\,[m] or $10^{-15}$\,[m].

The interactions of the muon neutrino in detectors are 
\begin{eqnarray}
& &\nu_{\mu}+e^{-} \rightarrow e^{-}+\nu_{\mu}, 
\label{numu-leptonic}\\
& &\nu_{\mu}+e^{-} \rightarrow \mu^{-}+\nu_{e}, 
\label{numu-lcharged}\\ 
& &\nu_{\mu}+A \rightarrow \mu^{-}+(A+1)+X, 
\label{numu-hcharged}\\ 
& &\nu_{\mu}+A \rightarrow \nu_{\mu} +A  +X,
\label{numu-hneutral}
\end{eqnarray}
hence the size of the neutrino wave packet  $\sqrt{\sigma_{\nu}}$ in processes $(\ref{numu-leptonic})$ and
$(\ref{numu-lcharged})$ is of order   $10^{-10}$,[m]
\begin{eqnarray}
\sqrt{\sigma_{\nu}} =10^{-10}~[\text{m}],
\label{mfp-neutrino-e}
\end{eqnarray}
 and the neutrino 
wave packet  $\sqrt{\sigma_{\nu}}$ in processes $(\ref{numu-hcharged})$ and
$(\ref{numu-hneutral})$ is of order   $10^{-15}$\,[m]
\begin{eqnarray}
\sqrt{\sigma_{\nu}} =10^{-15}~[\text{m}].
\label{mfp-neutrino-N}
\end{eqnarray}

The interactions of the electron neutrinos in detectors are 
\begin{eqnarray}
& &\nu_{e}+e^{-} \rightarrow e^{-}+\nu_{e}, 
\label{nue-leptonic}\\ 
& &\nu_{e}+A \rightarrow e^{-}+(A+1) +X,
\label{nue-hcharged}\\ 
& &\nu_{e}+A \rightarrow e  +A  +X.
\label{nue-hneutral}
\end{eqnarray}
The neutrino 
wave packet  $\sqrt{\sigma_{\nu}}$ in processes $(\ref{nue-leptonic})$
is of order   $10^{-10}$\,[m], Eq.\,$(\ref{mfp-neutrino-e})$, and the neutrino 
wave packet  $\sqrt{\sigma_{\nu}}$ in processes $(\ref{nue-hcharged})$ and
$(\ref{nue-hneutral})$ is of order   $10^{-15}$\,[m],
Eq.\,$(\ref{mfp-neutrino-N})$. They are treated in
the same way as the neutrino from the pion  decay.

From  
Eqs.\,$(\ref{mfp-neutrino-e})$ and $(\ref{mfp-neutrino-N})$, 
the neutrino have the wave packet sizes of the order $10^{-10}$\,[m]
or  $10^{-15}$\,[m].

We study  neutrinos described by the wave packets of these sizes in 
many particle processes. In this respect, the
neutrino wave packet of the present work is different from some  previous
works of wave packets that are connected with flavour neutrino
oscillations \cite{Kayser,Giunti,Nussinov,
Kiers,Stodolsky,Lipkin,Asahara}, where one particle properties of
neutrino at production are studied. It is important to study the
neutrino wave packet for our purpose of  studying the 
interference.
\subsection{Wave packet shape}
A particle of the finite coherence length is described by a wave packet, which
is  extended in the momentum and position around the centers $({\vec
p},{\vec X})$.
Although its precise shape is unknown generally in real
experiments, the physical quantity should have universal value that is
independent from the details of wave packet.  We 
shall obtain the physical quantity that has universal properties.
The quantity that 
depends on the wave packet shape is neither  genuine  nor 
universal and is not important. 

We give a summary of  the wave packets in the following, since the wave
packet may be unfamiliar to some readers. First, the wave 
packets are localized in the momentum and position around their centers \cite{Goldberger,newton,taylor,Sasakawa}.
The whole
set of the wave packets becomes complete set \cite{Ishikawa-Shimomura},
\begin{eqnarray}
\sum_{{\vec p},{\vec X}}|{\vec p},{\vec X},\beta \rangle
 \langle {\vec p},{\vec X}, \beta |=1,   
\label{complete-set}
\end{eqnarray}
independent from the shape parameter $\beta $ which is discussed later.

Second, the wave packet preserves the discrete symmetries such  as 
invariances under space  and time inversions, which have not been 
studied before but are  quite natural, since the origin of the wave 
packet is  the interactions of the
particle with matters in detectors and the wave packet sizes are 
estimated in the next section under this consideration. This interaction
has an origin  in quantum electrodynamics that preserves parity and 
time reversal symmetries. 
So the wave packets should preserve parity and time reversal invariances.
We study, hence, the wave packets which are   superpositions of the plane waves 
around the central momentum with  a weight function that has the same 
property under these transformations,
\begin{eqnarray}
\int d\vec{k}\, w({\vec k};{\vec p}\,) e^{i(Et-{\vec k}\cdot{\vec x})},
\end{eqnarray}
where the momentum ${\vec p}$ is the central value of the momentum.
In the present work, ${\vec k}$ is used for the integration variable and
${\vec p}$ is used for the central value of momentum.

Since the time reversal invariance is satisfied in the quantum
  electrodynamics  and weak
  interactions in the lowest order, the wave packet which is invariant 
under the time inversion is most important. Under the time inversion, 
the coordinate and momentum variables are transformed  into 
\begin{eqnarray}
& &{\vec x} \rightarrow {\vec x}, \nonumber\\
& &{\vec k}    \rightarrow -{\vec k},\, {\vec p}    \rightarrow -{\vec p}\nonumber\\
& &t \rightarrow -t,
\end{eqnarray}
and the plane wave is transformed  to its complex conjugate,
\begin{eqnarray}
 e^{i(Et-{\vec k}\cdot{\vec x})} \rightarrow e^{-i(Et-{\vec k}\cdot{\vec x})}= ({e^{i(Et-{\vec k}\cdot{\vec
 x})}})^{*}.
\end{eqnarray}
So when  the weight satisfies
\begin{eqnarray}
  w({\vec k};{\vec p}\, ) \rightarrow  (w(-{\vec k};-{\vec p}\, ))=
   (w({\vec k};{\vec p}\, ))^{*},
\label{time-reversal}
\end{eqnarray}
the state described by the wave packet is the superposition of the waves
which are transformed equivalently under the time inversion,
\begin{align}
\int d{\vec k}\,   w({\vec k};{\vec p}\, )e^{i(Et-{\vec k}\cdot{\vec x})} &\rightarrow \int
 d{\vec k}  \, w(-{\vec k};-{\vec p}\, )(e^{i(Et-{\vec k}\cdot{\vec
 x})})^{*}  \nonumber\\
&=\left(\int d{\vec k}\, 
 w({\vec k};{\vec p}\, )e^{i(Et-{\vec k}\cdot{\vec x})}\right)^{*}. 
\end{align}

Next we study the space inversion. Although neutrino violates
this symmetry, the wave packet may preserve.     Under the space
inversion, the coordinate and momentum variables are 
changed into 
\begin{eqnarray}
& &{\vec x} \rightarrow -{\vec x},\ {\vec k} \rightarrow -{\vec k},\ {\vec p} \rightarrow -{\vec p}, 
\end{eqnarray}
and the plane wave is changed into 
\begin{eqnarray}
e^{i(Et-{\vec k}\cdot{\vec x})} \rightarrow e^{i(Et-{\vec k}\cdot{\vec x})}.
\end{eqnarray}
So  when the weight satisfies 
\begin{eqnarray}
  w({\vec k};{\vec p}\,) \rightarrow  w(-{\vec k};-{\vec p}\,)= w({\vec k};{\vec p}\,),
\label{parity}
\end{eqnarray}
the state described by the wave packet is transformed in the following way,
\begin{align}
\int d{\vec k}\,   w({\vec k};{\vec p})e^{i(Et-{\vec k}\cdot{\vec x})} &\rightarrow \int
 d{\vec k}\,   w(-{\vec k};-{\vec p})e^{i(Et-{\vec k}\cdot{\vec x})}  \nonumber\\
&~=\int d{\vec k}\, 
 w({\vec k};{\vec p})e^{i(Et-{\vec k}\cdot{\vec x})},
\end{align}
in the same  way as the plane wave.

 The invariance under the time reversal, Eq.\,$(\ref{time-reversal})$, is
 a strong condition and leads 
the important result. We study wave packets 
that satisfy Eq.\,$(\ref{time-reversal})$. 
The simplest form of 
satisfying this property is the Gaussian
wave packet $|{\vec p},{\vec X},\beta_0 \rangle $
\begin{eqnarray}
|{\vec p},{\vec X},\beta_0 \rangle= \frac{N}{(2\pi)^{\frac{3}{2}}}\int d{\vec k}\, e^{-{\sigma \over 2}({\vec k}-{\vec p}\, )^2}e^{i\left(E({\vec k})(t-T)-{\vec
 k}\cdot({\vec x}-{\vec X})\right)}, 
\label{Gaussian-wp}
\end{eqnarray}  
where the parameter $\sigma$ shows the size of the wave packet in the
coordinate space and $N$ is the normalization factor.  We write a 
 wave packet of a parameter $\beta$ as $|{\vec p},{\vec
X},\beta \rangle$. The Gaussian wave
packet is the state  at $\beta = \beta_0$ and  non-Gaussian wave
packets are the states at $\beta \neq \beta_0$.
The calculations
using the Gaussian wave packet are presented in most places hereafter 
for the sake of simplicity, except the derivation of the large distance
behavior of the neutrino wave function. It has the universal form 
regardless of the details of the wave packets, as far as the invariance
under the time inversion holds. 
The theorem for the general wave packets that are invariant under the time
inversion  Eq.\,$(\ref{time-reversal})$ holds and is given in 
Section 4.  It is worthwhile to verify the results using explicit 
calculations  of  the  Gaussian wave packet or  various
non-Gaussian wave packets, which are made partly in the text and  
in the appendix.

The normal physical quantity of microscopic physics is obtained from 
ordinary scattering which has no dependence upon the distance or time
interval between the initial and final states. 
This is because the length of the microscopic system is so small that 
the size of experimental apparatus is regarded infinite and  the boundary
conditions of ordinary scatterings  which are defined  at $t=\pm \infty$ 
of plane waves are suitable.  Hence  the  amplitude and probability are
defined using  plane waves and the boundary conditions at 
$t=\pm \infty$ ensure the  independence of the probability from the 
distance and    particle's coherence length. 

For the neutrino the situation is different because the neutrino mass is
so small that a new energy  scale defined by ${m_\nu^2/E_{\nu}}$
becomes extremely small and a spatial length which is inversely
proportional to this energy becomes macroscopic length.  It is
important to find a physical quantity that depends on this length,
since it may give a new important information.
To find a scattering amplitude and probability that have a dependence
on the length or time interval, we use   the amplitude defined from 
wave packets of  finite spatial sizes, $\sigma$. Those  wave packets 
that have   central positions  and
are  localized well around the central positions and  that have  the
same properties  in the momentum variable are suitable for this  purpose.  
The simplest wave packets of having these properties are Gaussian wave
packets  Eq.\,$(\ref{Gaussian-wp})$ which satisfy the minimum uncertainty 
relation between the 
variances of the position and momentum. 
Non-Gaussian  wave
packet parameterized by a parameter $\beta $ and has  larger 
uncertainties  is   easily defined by  multiplying Hermitian polynomials 
to the Gaussian function.  

From Eq.\,$(\ref{complete-set})$, the total transition
probability is
independent from the parameter $\beta$
\begin{align}
\label{completeness-total}
&\sum_{{\vec p}_{\nu},{{\vec X}_{\nu}}} |\langle neutrino;{\vec p}_{\nu},{\vec X}_{\nu} 
 ,\beta;muon |T | pion;{\vec p}_{\pi},{\vec X}_{\pi},T_{\pi}\rangle |^2 \nonumber\\
&=\sum_{{\vec p}_{\nu},{{\vec X}_{\nu}}} |\langle neutrino;{\vec p}_{\nu},{\vec X}_{\nu} 
 ,\beta_0;muon |T | pion;{\vec p}_{\pi},{\vec X}_{\pi},T_{\pi}\rangle |^2,
\end{align}
which also agree to that of the plane waves. The probability of the
finite distance is defined by restricting  the center positions ${\vec
X}$ in the 
inside of a finite  spatial region $V$. The completeness condition 
for the function in this region $V$ is then 
 \begin{eqnarray}
\label{completeness-finite-volume}
\sum_{{\vec p},{\vec X}<V}|{\vec p},{\vec X},\beta_0 \rangle
 \langle {\vec p},{\vec X}, \beta_0 |=   \sum_{{\vec p},{\vec X}<V}|{\vec p},{\vec X},\beta \rangle
 \langle {\vec p},{\vec X}, \beta |,
\end{eqnarray}  
and  the probability satisfies
\begin{align}
\label{completeness-finite-volume2}
&\sum_{{\vec p}_{\nu},{{\vec X}_{\nu}<V}} |\langle neutrino;{\vec p}_{\nu},{\vec X}_{\nu} 
 ,\beta;muon |T | pion;{\vec p}_{\pi},{\vec X}_{\pi},T_{\pi}\rangle |^2  \nonumber\\
&=\sum_{{\vec p}_{\nu},{{\vec X}_{\nu}<V}} |\langle neutrino;{\vec p}_{\nu},{\vec X}_{\nu} 
 ,\beta_0;muon |T | pion;{\vec p}_{\pi},{\vec X}_{\pi},T_{\pi}\rangle |^2,
\end{align}
and is also independent from the $\beta$.  
We study the probability that is the average over a finite neutrino energy 
region $V_p$ 
\begin{align}
 \label{completeness-average momentum}
&\sum_{{{\vec p}_{\nu}<V_p},{{\vec X}_{\nu}<V}} |\langle neutrino;{\vec p}_{\nu},{\vec X}_{\nu} 
 ,\beta;muon |T | pion;{\vec p}_{\pi},{\vec X}_{\pi},T_{\pi}\rangle |^2,  
\end{align}
and extract the physical quantity.  For the genuine physical quantity, 
the value should be determined uniquely. 
The energy uncertainty of the
neutrino experiment is of the order 10 percent of the total 
neutrino energy. This magnitude  is the same order as that of the minimum 
uncertainty if the neutrino energy is $1$ [GeV].  So the minimum wave
packet is applied directly in this case. The probability for 
a  larger energy uncertainty 
 is  computed using  the probability of the small energy 
uncertainty. The dependence of the neutrino probability on the distance 
will be shown  to have the universal behavior. 
  
\section{Position-dependent amplitude of neutrino }
Using the wave packet representation obtained in the previous section, we find  
the  position-dependent neutrino amplitude.    
\subsection{Semileptonic decay of the pion }
\subsubsection{Weak interaction}
Semileptonic decay of a pion is described by the weak Hamiltonian  
\begin{eqnarray}
& &H_{w}=g \int d{\vec x}\, {\partial_{\mu}
 }\varphi(x)J_{V-A}^{\mu}(x)=-igm_{\mu}\int d{\vec x} \, \varphi(x)J_5(x),
\label{weak-hamiltonian}\nonumber\\
& &J_{V-A}^{\mu}(x)=\bar
 \mu(x)\gamma^{\mu}(1-\gamma_5)\nu(x),J_5(x)=\bar
 \mu(x)(1-\gamma_5)\nu(x),
\end{eqnarray}
where ${\varphi(x)}$, $\mu(x)$, and $\nu(x)$ are the pion field, muon
field, and neutrino field \cite{J-J}. In the above equations, 
$g$ is the coupling
strength, $J_{V-A}^{\mu}(x)$, and $J_5(x)$ are the 
leptonic charged vector current, and leptonic  pseudoscalar. The
coupling is expressed with Fermi coupling $G_F$ and pion decay constant $f_\pi$,
\begin{eqnarray} 
g={G_F \over \sqrt 2}f_{\pi}.
\end{eqnarray}
\subsubsection{Neutrino production amplitude in the pion decay}
When  the number of the neutrino events is large enough, the
 distributions of the neutrino events are computed with the quantum
mechanical wave functions.  For the electron bi-prism experiments of
Tonomura et al \cite{tonomura}, the electrons' interferences of the
sharp energy become
visible when the
total number becomes significant. Even though the initial electrons have no
 correlation and  each event occurs randomly,  when number of the 
electrons becomes large, the signal  exceeds the statistical fluctuation
 and  the distribution shows the
 interference behavior. A large number of events 
is also necessary for the quantum mechanical interferences become visible for 
the neutrinos from the pion decays. In this situation each pion decay
 occurs randomly and its distribution is computed using quantum
 mechanical wave functions. We study this situation hereafter and 
compute the amplitude and probability quantum  mechanically.  
    
A wave function which describes a pion and its decay products satisfies
a Schr\"{o}dinger equation
\begin{eqnarray}
i\hbar {\partial \over \partial t}|\Psi(t)\rangle=(H_0+H_w)
 |\Psi(t)\rangle,
\end{eqnarray}
where $H_0$ stands for the free Hamiltonian and $H_w$ stands for the
interaction Hamiltonian Eq.\,(\ref{weak-hamiltonian}). The solution is 
\begin{eqnarray}
& &|\Psi(t) \rangle= |\text {pion} (t)\rangle+|\text {muon,~neutrino}(t)
 \rangle\label{state-vector}, \nonumber\\
& &|\text {muon,~neutrino}(t) \rangle=\int_{t_0}^{t} {d t' \over i\hbar}H_w(t') |\text{pion}
 (t')\rangle,
\end{eqnarray}
in the first order of $H_w$. A muon and a neutrino are described by one
wave function of Eq.\,$(\ref{state-vector} )$, $|\text {muon,~neutrino}(t)
\rangle$, and all the informations of this system are computed from this
wave function. Physical quantities were
studied with this wave function \cite{Asahara} and the
finite time interval effect was found.

Here the matrix elements between $|\text {muon,~neutrino}(t)
\rangle$ and the state $|\mu ,\nu \rangle $ defined next is studied
instead of the wave function itself.  Since all the informations of the
 wave function $|\text {muon,~neutrino}(t)
\rangle$ are included in this matrix element, as far as a
complete set of $|\mu ,\nu \rangle $ is used,  this matrix
element is equivalent to the wave function. Furthermore   
the position-dependent informations on the decay process
are found by the use of the wave packets. The coherent
phenomenon of the neutrino which depends on  the position
and the neutrino detection probability are  studied  here with 
this matrix element.

The transition amplitude from  a pion located around the position ${\vec
X}_\pi$ to a neutrino at the position ${\vec
X}_{\nu}$ and the muon is studied hereafter. 
Because the muon is not observed and all the states are
summed over, the most convenient wave is used for the muon. This amplitude is  
\begin{eqnarray}
T=\int d^4x \, \langle {\mu},{\nu}   |H_{w}(x)| \pi \rangle,
\end{eqnarray}
where the states are described by  the wave packets 
of central values of momenta and  coordinates and the widths
in the form   
\begin{eqnarray}
|\pi \rangle=   | {\vec p}_{\pi},{\vec X}_{\pi},T_{\pi}  \rangle,\ 
|\mu ,\nu \rangle=   |\mu,{\vec p}_{\mu};\nu,{\vec p}_{\nu},{\vec X}_{\nu},T_{\nu}          \rangle.
\end{eqnarray}
The states are defined by the matrix elements, 
\begin{align}
&\langle 0|\varphi(x)|{\vec p}_{\pi},{\vec X}_{\pi},T_{\pi}  \rangle
= N_{\pi}\rho_{\pi}\int d{\vec k_\pi} \, e^{-{\sigma_{\pi} \over 2}({\vec
 k}_\pi-{\vec p}_{\pi})^2}e^{-i\left(E({\vec
 k}_\pi)(t-T_{\pi}) - i{\vec
 k}_\pi\cdot({\vec x}-{\vec X}_{\pi})\right)}
\nonumber \\
&\approx 
N_{\pi}\rho_{\pi}\left({2\pi \over \sigma_{\pi}}\right)^{\frac{3}{2}}e^{-{1 \over 2 \sigma_{\pi} }\left({\vec x}-{\vec X_{\pi}}-{\vec v}_{\pi}(t-T_{\pi})\right)^2}
e^{-i\left(E({\vec p}_{\pi})(t-T_{\pi}) - i{\vec
 p}_{\pi}\cdot({\vec x}-{\vec X}_{\pi})\right)} ,
\label{pion-wf}\\
&\langle \mu,{\vec p}_{\mu};\nu,{\vec p}_{\nu},{\vec X}_{\nu},T_{\nu}|\bar
 \mu(x) (1 - \gamma_5) \nu(x) |0 \rangle \nonumber\\
&= \frac{N_{\nu}}{(2\pi)^{{3}}}\int d{\vec k}_{\nu}e^{-{\sigma_{\nu} \over 2}({\vec k}_{\nu}-{\vec p}_{\nu})^2}\left({m_{\mu}
 \over E({\vec p}_{\mu})}\right)^{\frac{1}{2}}\left({m_{\nu} \over E({\vec
 k}_{\nu})}\right)^{\frac{1}{2}}  \bar u({\vec p}_{\mu}) (1 -\gamma_5) \nu
 ({\vec k}_{\nu})\nonumber\\
 &\times e^{i\left(E({\vec p}_{\mu})t-{\vec
 p}_{\mu}\cdot{\vec x}\right)}e^{i\left(E({\vec k}_{\nu})(t-T_{\nu})-{\vec
 k}_{\nu}\cdot({\vec x}-{\vec X}_{\nu})\right)},\label{mu-nu-wf}
\end{align}
where
\begin{align}
& N_{\pi} = \left(\frac{\sigma_{\pi}}{\pi}\right)^{\frac{3}{4}},~N_{\nu}
 = \left(\frac{\sigma_{\nu}}{\pi}\right)^{\frac{3}{4}},\rho_{\pi}=
\left(\frac{1}{2E_{\pi}(2\pi)^3}\right)^{\frac{1}{2}}.
\end{align}
In this paper the spinor's normalization is
\begin{eqnarray}
\sum_{s} u(p,s)\bar u(p,s)={\gamma \cdot p+m \over 2m}.
\end{eqnarray}
In the above equation  the pion's life time is ignored.
The sizes, $\sigma_{\pi}$ and
$\sigma_{\nu}$,  in Eqs.\,$(\ref{pion-wf})$ and 
  $(\ref{mu-nu-wf})$   are sizes of the pion wave packet and  of the 
neutrino wave packet. Minimum wave packets are used in most of the
present  paper but non-minimum wave packets are studied and it is shown
that  our results are the same.
\footnote{For the non-minimal wave packets which  have larger uncertainties 
Hermitian polynomials of ${\vec k}_{\nu}-{\vec p}_{\nu}$ are multiplied to
the right-hand side of
Eq.\,$(\ref{mu-nu-wf})$ . The completeness of the wave packet states is
also satisfied for the
non-minimum case and the total probability and the probability of the
finite distance and time are  the same. We will confirm in the text and 
appendix that the universal long-range correlation of the 
present work is independent from the wave 
packet shape as far as the wave packet is invariant under the time inversions.}

From the results of the previous section, the size of pion wave packet is of the 
order $0.5-1.0$\,[m] and  the momentum has a
small width. So the pion momentum  is integrated easily, and is 
replaced with  its  central value ${\vec p}_\pi$ and the final 
expression  of Eq.\,$(\ref{pion-wf})$ is obtained. 
For the neutrino, the size of wave packet should be the size of minimum 
physical system  that a neutrino interacts, i.e., the nucleus. Hence to study
neutrino interferences, we use the  nuclear size for $\sigma_{\nu}$. 

The
amplitude $T$ for  one pion to decay into a neutrino and a muon  is
written in the form
\begin{eqnarray}
T =  \int d t d{\vec x} T(t,{\vec x}),T(t,{\vec x})=\int  d{\vec k}_{\nu}T(t,{\vec x},k_{\nu})\label{amplitude}
\end{eqnarray}
where
\begin{eqnarray}
 T(t,{\vec x},k_{\nu}) &=&ig m_{\mu} N'' 
e^{-{1 \over 2 \sigma_{\pi} }\left({\vec x}-{\vec X_{\pi}}-{\vec v}_{\pi}(t-T_{\pi})\right)^2}
e^{-i\left(E({\vec p}_{\pi})(t-T_{\pi}) - i{\vec
 p}_{\pi}\cdot({\vec x}-{\vec X}_{\pi})\right)} \nonumber \\
 & &\times e^{i\left(E(\vec{p}_{\mu})t-{\vec p}_{\mu}\cdot{\vec
		x}\right)} \bar{u}({\vec p}_{\mu}) (1 - \gamma_5)
 \nu({\vec k}_{\nu})e^{i\left(E({\vec k}_{\nu})(t-T_{\nu})-{\vec
			 k}_{\nu}\cdot({\vec x}-{\vec X}_{\nu})\right)
 -\frac{\sigma_{\nu}}{2}({\vec k}_{\nu}-{\vec p}_{\nu})^2}, \nonumber \\
 N''&=&N_{\pi}N_{\nu}\left({2\pi \over \sigma_{\pi}}\right)^{\frac{3}{2}}N_0,~N_0= \left(\frac{m_{\mu}m_{\nu}}{
 E_{\mu}E_{\nu}}\right)^{\frac{1}{2}}\times\frac{\rho_{\pi}}{(2\pi)^3}.
\end{eqnarray}
This amplitude depends upon the coordinates $(t,{\vec x})$ explicitly 
and is  not invariant  under the translation.  
         

\subsection{Neutrino wave function }
We integrate the neutrino momentum of Eq.\,$(\ref{amplitude})$ 
by applying Gaussian integral and find the coordinate representation of
the neutrino wave function. It is found that 
a phase of neutrino wave function has a particular form that is 
proportional to the square of the mass and inversely proportional to 
the neutrino energy.  


\subsubsection{Phase of neutrino wave function}
For not so large $t-T_{\nu}$ region, ${\vec k}_{\nu}$ is integrated 
around the central momentum ${\vec p}_{\nu}$ in
Eq.\,$(\ref{amplitude})$. 
The amplitude becomes,
 \begin{eqnarray}
T(t,{\vec x})&=&igm_{\mu}\tilde N   e^{-{1 \over 2 \sigma_{\pi} }\left({\vec x}-{\vec X_{\pi}}-{\vec v}_{\pi}(t-T_{\pi})\right)^2}
e^{-i\left(E({\vec p}_{\pi})(t-T_{\pi}) - i{\vec
 p}_{\pi}\cdot({\vec x}-{\vec X}_{\pi})\right)}
e^{iE({\vec p}_{\mu})t-i{\vec
 p}_{\mu}\cdot{\vec x}}\nonumber\\
& &\times\bar u({\vec p}_{\mu}) (1 - \gamma_5) \nu ({\vec p}_{\nu})
e^{i\phi(x)}  e^{-{1 \over 2\sigma_{\nu} }({\vec x}-{\vec X}_{\nu} -{\vec
v}_{\nu}(t-T_{\nu}))^2},
\label{neutrino-position-amplitude}
\end{eqnarray}
where $\tilde N$ is the normalization factor, ${\vec v}_{\nu}$ is the neutrino velocity, and $\phi$ is the phase of neutrino
wave function. They are given by  
\begin{eqnarray}
& &\tilde N=N_{\pi} N_{\nu}\left(\frac{2\pi}{\sigma_{\pi}}\right)^{\frac{3}{2}}\left(\frac{2\pi}{\sigma_{\nu}}\right)^{\frac{3}{2}}N_0,\\
& &\phi(x)=E({\vec
 p}_{\nu})(t-T_{\nu})-{\vec  p}_{\nu}\!\cdot\!({\vec x}-{\vec X}_{\nu}).
\label{neutrino-phase} 
\end{eqnarray}
The phase  $\phi(x)$ is rewritten
for a small wave packet by
substituting the central value ${\vec x}$ of neutrino's Gaussian
function 
\begin{eqnarray}
{\vec x}={\vec X}_{\nu}+{\vec v}_{\nu}(t-T_{\nu}),
\label{center-coordinate}
\end{eqnarray}
 in the form   
\begin{eqnarray}
\bar \phi_g(t-T_{\nu})
\label{phase}
&=&E({\vec p}_{\nu})(t-T_{\nu})-{\vec p}_{\nu}\cdot{\vec
 v}_{\nu}(t-T_{\nu})\nonumber\\
&=&{E^{2}_{\nu}(\vec p_{\nu})-{\vec p}_{\nu}^{~2} \over E_{\nu}({\vec p}_{\nu})}(t-T_{\nu})= {m_{\nu}^2 \over E_{\nu}(\vec{p}_{\nu})} (t-T_{\nu}),
\end{eqnarray}
which  has a typical form of the relativistic particle. The phase
becomes  proportional to the neutrino mass squared and inversely proportional to
the neutrino energy.   The magnitude of the phase at the position
Eq.\,$(\ref{center-coordinate})$ is small due to the cancellation between
the oscillation in time and space. This cancellation does
not occur in the derivatives of the phase with respect to the
coordinate, which  are given in the form
\begin{eqnarray}
\frac{\partial}{\partial x_{\mu}} \phi=p_{\nu}^{\mu}. 
\label{derivative-phase}
\end{eqnarray}
This is not  proportional to the square of neutrino mass but is  
determined by the energy and momentum.

When the position is moving with the light velocity
\begin{eqnarray}
{\vec x}={\vec X}_{\nu}+{\vec c}(t-T_{\nu}),\ |{\vec c}\,|=1,
\end{eqnarray}
then the phase  is given by
\begin{eqnarray}
\bar \phi_c(t - T_\nu)=E({\vec p}_{\nu})(t-T_{\nu})-{\vec p}_{\nu}\!\cdot{\vec
 c}\,(t-T_{\nu})
=\frac{m_{\nu}^2}{2 E_{\nu}(\vec{p}_{\nu})} (t-T_{\nu}),\label{light-phase}
\end{eqnarray}
and becomes a half of $\bar{\phi_g}$.

Interference of the neutrino due to this slow phase of the energy
$E_{\nu}$ and $E_{\nu}+\Delta E_{\nu}$ is identical 
if the  neutrino phase $\bar \phi_{c}$ satisfies
\begin{eqnarray}
\Delta \bar \phi_c=\delta
 E_{\nu} {\partial  \over \partial E_{\nu}} \bar \phi_c  \ll \pi.
\end{eqnarray}
Hence the neutrinos of the energy width $\Delta E_{\nu}$ show the same
interference.    
We will see later that this is actually satisfied for the light neutrino
of the energy width around $100$\,[MeV].

The slow phase $\bar{\phi_c}$ of Eq.\,$(\ref{light-phase})$ is a characteristic 
feature of
the neutrino, and  is
   the intrinsic property of the neutrino wave function along the light
   cone. Using the
   Gaussian wave packet, this phase has been obtained easily. Since this 
phase is 
    genuine of the neutrino wave function, the same behavior is 
obtained in more general cases
   including the spreading wave packet or  non-Gaussian wave packets.     
   The spreading of the wave packet  is
negligible in the longitudinal direction and is not so small in the
transverse direction. Its effect has been  ignored for simplicity in
   this section  and will be studied in the latter section and
   appendix. It will be shown there that the spreading of the  wave
   packet in the transverse 
direction modifies the ${\vec k}_{\nu}$ integration but the final result 
turns actually into the same. For  general wave packets,
 if the completeness, 
Eq.\,$(\ref{completeness-finite-volume})$ 
is applied, the total probability satisfies  Eq.\,$(\ref{completeness-finite-volume2})$.
  The transition amplitude in this  case has the same phase $\phi(x)$,
  Eq.\,$(\ref{neutrino-phase})$ and is
  written in the
  form,
 \begin{eqnarray}
T(t,{\vec x})&=&igm_{\mu}\tilde N e^{-{1 \over 2 \sigma_{\pi} }\left({\vec x}-{\vec X_{\pi}}-{\vec v}_{\pi}(t-T_{\pi})\right)^2}
e^{-i\left(E({\vec p}_{\pi})(t-T_{\pi}) - i{\vec
 p}_{\pi}\cdot({\vec x}-{\vec X}_{\pi})\right)}
e^{i(E({\vec p}_{\mu})t-{\vec
 p}_{\mu}\cdot{\vec x})}\nonumber\\
&\times&\bar u({\vec p}_{\mu}) (1 - \gamma_5) \nu ({\vec p}_{\nu})
e^{i\phi(x)} w\left({\vec x}-{\vec X}_{\nu} -{\vec
v}_{\nu}(t-T_{\nu})\right) ,
\label{neutrino-position-gamplitudeG}
\end{eqnarray} 
where $w({\vec x})$ is the amplitude of neutrino wave function obtained
by the Fourier transformation from the $w({\vec p},{\vec k})$. The
spreading effect is negligible for the small time interval and $w({\vec
x})$ is given in the form
\begin{eqnarray}
\label{fourier-transform1}
w({\vec x})=\int d\tilde{\vec{k}}e^{i\tilde{\vec{k}}\cdot{\vec
 x}}w({\tilde {\vec k}}),~\tilde {\vec k}={\vec k}-{\vec p},
\end{eqnarray}
and 
decreases fast
with   $|{\vec x}|$.  Moreover from  Eq.\,$(\ref{time-reversal})$, this 
function is real function
\begin{eqnarray}
w({\vec x})=w({\vec x})^{*},
\label{real-even-wavefunction}
\end{eqnarray}
which leads the universal long-distance behavior to the neutrino probability.
The spreading effect is included when the long-range component of the
correlation is studied in Section 4.

\subsubsection{Position-dependence,  energy momentum
   non-conservation, and interferences}
When the space-time coordinates $(t,{\vec x})$ are 
integrated in the amplitude of the plane waves, the delta function 
of the energy and momentum conservation emerges.  The scattering
amplitude  with this 
delta function shows that the final states have the same energy and momentum
with the initial state. On the other hand, the position-dependent
amplitude $T$ for the wave packets is not invariant under the 
translation and has no delta function. So the  energy and momentum of 
the final state are not necessary the same as the initial state. 
The states which do not satisfy  the energy and momentum conservation 
should be  included to get consistent results from the completeness.  
This amplitude  shows the position-dependent behavior, from which
a new information is found.

Let us compare the neutrino velocity with the
light velocity. The
 neutrino of energy  $1~[\text{GeV}]$ and  the mass 
$1~[\text{eV}/c^2]$    has a velocity
\begin{eqnarray}
v/c=1-2\epsilon, \ \epsilon=\left({m_{\nu}c^2 \over E_{\nu}}\right)^2=5\times 10^{-19},
\end{eqnarray} 
hence the neutrino propagates the  distance $l$, where  
\begin{eqnarray}
l=l_0(1-\epsilon)=l_0-\delta l,\delta l= l_0\times \epsilon,
\end{eqnarray}
while  the light propagates  the distance $l_0$. This difference of distance,
$\delta l$, becomes
\begin{eqnarray}
& &\delta l=5\times 10^{-17}~[\text{m}]; ~l_0=100~[\text{m}], \\ 
\label{neutrino-ovelapp}
& &\delta l=5\times 10^{-16}~[\text{m}]; ~l_0=1000~[\text{m}] ,
\end{eqnarray}
which are much smaller than the sizes of the above wave packets
Eq.~$(\ref{mfp-pion})$.
This gives the important
effect for the neutrino amplitude at the nuclear target or the atom
target to show interference. The geometry of the neutrino interference 
is shown in
Fig.~\ref{fig:geo}. The neutrino wave produced at a time $t_1$ arrives
to one nucleus or atom in the detector and is added to the wave
produced  at $t_2$ and arrives to the same    nucleus or atom same
time. A constructive interference of waves is shown in the text.


\section{Position-dependent  probability and interference }
The probability of detecting neutrino at a finite distance 
is studied  in this section. Particularly the finite-time interval T
correction, i.e., the deviation of the transition probability from the
T-linear form, is obtained. For this purpose, the total probability is 
expressed with the  two point correlation  function of decay process, 
which has the light-cone singularity. The light-cone singularity leads  
the finite-distance correction of universal form.

\subsection{Transition probability }
The transition probability of a pion of momentum ${\vec
p}_{\pi}$ located at  space-time position $(T_{\pi},{\vec
X}_{\pi})$  to a neutrino of momentum ${\vec p}_{\nu}$ at  space-time 
position $(T_{\nu},{\vec X}_{\nu})$ and a muon of momentum ${\vec p}_{\mu}$ is written in the form 
\begin{eqnarray}
|T|^2 &=& g^2 m_{\mu}^2 
|\tilde N|^2 \int d^4x_1 d^4x_2
S_{5}(s_1,s_2)
\nonumber\\
 &\times& e^{i( \phi(x_1) -\phi(x_2))}  
e^{-{1 \over 2\sigma_{\nu} }\left({\vec x}_1-{\vec X}_{\nu} -{\vec
			     v}_{\nu}(t_1 - T_{\nu})\right)^2}e^{-{1 \over 2\sigma_{\nu} }\left({\vec x}_2-{\vec X}_{\nu} -{\vec v}_{\nu}(t_2-T_{\nu})\right)^2}
\nonumber \\
&\times& e^{-i\left(E({\vec p}_{\pi})(t_1 - T_{\pi})-{\vec p}_{\pi}\cdot({\vec x}_1-{\vec
X}_{\pi})\right)}e^{i\left(E({\vec p}_{\pi})(t_2 - T_{\pi})-{\vec p}_{\pi}\cdot({\vec x}_2-{\vec
X}_{\pi})\right)}\nonumber \\
&\times&e^{i\left(E({\vec p}_{\mu})t_1-{\vec p}_{\mu}\cdot{\vec x}_1\right)}
 e^{-i\left(E({\vec p}_{\mu})t_2-{\vec p}_{\mu}\cdot{\vec x}_2\right)}
\nonumber \\
&\times&e^{-{1 \over 2 \sigma_{\pi} }\left({\vec x}_1-{\vec X_{\pi}}-{\vec
 v}_{\pi}(t_1-T_{\pi})\right)^2}e^{-{1 \over 2 \sigma_{\pi} }\left({\vec x}_2-{\vec
 X_{\pi}}-{\vec v}_{\pi}(t_2-T_{\pi})\right)^2},  \label{probability}
\end{eqnarray}
where $S_{5}(s_1,s_2)$ stands for the products of Dirac
spinors and their  complex conjugates,   
\begin{eqnarray}
S_{5}(s_1,s_2)=\left(\bar u({\vec p}_{\mu})
 (1 - \gamma_5) \nu ({{\vec p}_{\nu}})\right)\left(\bar u({\vec p}_{\mu})
 (1 -  \gamma_5) \nu ({{\vec p}_{\nu}})\right)^{*},
\label{spinor-1}
\end{eqnarray}
and its spin summation is  given by
\begin{eqnarray}
S^{5}&=&\sum_{s_1,s_2}S^{5}(s_1,s_2)
=\frac{2}{m_{\nu}m_{\mu}}(p_{\mu}\!\cdot\! p_{\nu}).\label{spinor-2}
\end{eqnarray}

For the plane waves, the coordinates $x_1$ and $x_2$ are integrated
easily and
$(\delta^{(4)}(p_{f}-p_i))^2=\delta^{(4)}(p_{f}-p_i)\delta^{(4)}(0)$ is   
obtained. Thus the four dimensional momentum is conserved and the
probability is proportional to the volume $V$, which is cancelled with 
the normalization factor and
T from $\delta^{(4)}(0)=V \text T$. The probability is proportional to
T and the proportional constant gives the decay rate per time. Here
we calculate the finite-time interval correction for the neutrino
detection process. The neutrino is observed through the interaction with
the nucleus and is expressed with the wave packet. In order to obtain
the finite-time interval correction, we write  
the probability with the 
correlation function of the pion and muon system. The transition
probability which is integrated in momenta of the final state and
average over the initial momentum is
written, then, in the form 
 \begin{align}
&\int d{\vec p}_{\pi}\rho_{exp}({\vec p}_{\pi}) d{\vec
 p}_{\mu}d{\vec p}_{\nu}  \sum_{s_1,s_2}|T|^2 
 \label{probability-correlation}\nonumber \\
&= g^2 m_{\mu}^2 |N_{\pi\nu}|^2\frac{2}{(2\pi)^3}\int d{\vec p}_{\nu} \rho_{\nu}^2 d^4x_1 d^4x_2
e^{-{1 \over 2\sigma_{\nu} }\left({\vec x}_1-{\vec X}_{\nu} -{\vec
v}_\nu(t_1-T_{\nu})\right)^2}e^{-{1 \over 2\sigma_{\nu} }\left({\vec x}_2-{\vec X}_{\nu} -{\vec v}_\nu(t_2-T_{\nu})\right)^2}\nonumber \\
&  \times 
\Delta_{\pi,\mu}(\delta t,\delta {\vec x})
e^{i \phi(\delta x_{\mu})} 
e^{-{1 \over 2 \sigma_{\pi} }\left({\vec x}_1-{\vec X_{\pi}}-\bar{\vec
 v}_{\pi}(t_1-T_{\pi})\right)^2} e^{-{1 \over 2 \sigma_{\pi} }\left({\vec x}_2-{\vec
 X_{\pi}}-\bar {\vec v}_{\pi}(t_2-T_{\pi})\right)^2}, \nonumber \\
&
N_{\pi\nu} =
\left(\frac{4\pi}{\sigma_{\pi}}\right)^{\frac{3}{4}}\left(\frac{4\pi}{\sigma_{\nu}}\right)^{\frac{3}{4}},~\rho_{\nu}=\left(\frac{1}{2E_{\nu} (2\pi)^3 }\right)^{\frac{1}{2}},~\delta  x= x_1-x_2,
\end{align}
using the correlation function $\Delta_{\pi,\mu}(\delta t,\delta {\vec x})$. The correlation
function is defined with   a pion's momentum distribution $\rho_{exp}({\vec p}_{\pi})$, by
\begin{align}
\Delta_{\pi,\mu} (\delta t,\delta {\vec x})=
 {\frac{1}{(2\pi)^3}}\int
{d {\vec p}_{\pi} \over E({\vec p}_{\pi})}\rho_{exp}({\vec p}_{\pi})
{d {\vec p}_{\mu} \over E({\vec p}_{\mu})}  (p_{\mu}\!\cdot\! p_{\nu})
 e^{-i\left(\{E({\vec
 p}_{\pi})-E({\vec p}_{\mu})\}\delta t-({\vec p}_{\pi}-{\vec
 p}_{\mu})\cdot \delta {\vec x})\right)}.
\label{pi-mucorrelation}
\end{align} 

 In the above equation, 
Eqs.\,($\ref{phase}$) and ($\ref{derivative-phase}$) were 
substituted.  
The muon momentum is integrated in whole positive
energy  region, because the muon is not observed. If the muon is
observed and its momentum is measured together  with the neutrino, then 
the muon momentum is integrated in the finite-energy region as in the
inclusive hadron reactions \cite{mueller}.  Pion
momentum is integrated in the region specified by the initial pion beam.
The velocity ${\vec v}_{\pi} $ in the pion Gaussian factor was
replaced with its average $\bar {\vec v}_{\pi}$. This is verified from
the large spatial size of the pion wave packet discussed in the 
previous section. 
If   the pion's momentum is fixed to one value, the correlation function 
 \begin{align}
\tilde \Delta_{\pi,\mu} (\delta t,\delta {\vec x})=
 {\frac{1}{(2\pi)^3}}
{1 \over E({\vec p}_{\pi})}
\int {d {\vec p}_{\mu} \over E({\vec p}_{\mu})}  (p_{\mu}\!\cdot\! p_{\nu})
 e^{-i\left(\{E({\vec
 p}_{\pi})-E({\vec p}_{\mu})\}\delta t-({\vec p}_{\pi}-{\vec
 p}_{\mu})\cdot \delta {\vec x}\right)}
\label{pi-mucorrelation2}
\end{align} 
is used instead of Eq.$( \ref{pi-mucorrelation})$.    
\subsection{Light-cone singularity   }
The correlation function $\tilde \Delta_{\pi,\mu}(\delta t,\delta {\vec
x})$  has a singularity near 
the light-cone region
\begin{align}
\lambda=\delta t^2-{\left|\delta\vec x\right|}^2 = 0,
\end{align}
which is extended into a large  $|\delta {\vec x}|$ and is independent
from ${\vec p}_{\pi}$. So $ \Delta_{\pi,\mu}(\delta t,\delta {\vec
x})$ has the same light-cone singularity. 
So the probability Eq.\,$(\ref{probability-correlation})$ has a 
large finite-distance 
correction due to the singularity of $\Delta_{\pi,\mu}
(\delta t,\delta {\vec x})$.  
We find the
light-cone singularity of    $\tilde \Delta_{\pi,\mu}
(\delta t,\delta {\vec x})$ \cite{Wilson-OPE}  in the following 
and apply the result also to $\Delta_{\pi,\mu}
(\delta t,\delta {\vec x})$.

\subsubsection{ Separation of singularity  }

If the particles are plane waves, the  energy and momentum 
are strictly conserved and the momenta satisfy  
\begin{eqnarray}
p_{\pi}=p_{\mu}+p_{\nu},\ 
(p_{\pi}-p_{\mu})^2=m_{\nu}^2 \approx 0.
\label{lightlike}
\end{eqnarray}
Hence the momentum difference $p_{\pi}-p_{\mu}$ is almost
on the light cone and the $\Delta_{\pi,\mu} (\delta
t,\delta {\vec x})$ around  the light cone,  $\lambda=0$, is important
in  Eq.$(\ref{probability-correlation})$. 
 In order to extract the singular term from  $\tilde \Delta_{\pi,\mu} (\delta
t,\delta {\vec x})$, we write  the integral in the form   
\begin{align}
&\tilde \Delta_{\pi,\mu} (\delta t,\delta {\vec x})=
 {\frac{1}{(2\pi)^3}} {1 \over E({\vec
 p}_{\pi})}I(p_{\pi},\delta x),\nonumber\\
&I(p_{\pi},\delta x)={2 \over \pi} \int d^{(4)}p_{\mu} \, \theta(p_{\mu}^0)
(p_{\mu}\!\cdot\! p_{\nu}) \text {Im}\left[1 \over p_{\mu}^2-m_{\mu}^2-i\epsilon\right]
 e^{-i\left(\{E({\vec
 p}_{\pi})-E({\vec p}_{\mu})\}\delta t-({\vec p}_{\pi}-{\vec
 p}_{\mu})\cdot \delta {\vec x}\right)}. 
\end{align}
Next the integration variable is changed from $p_{\mu}$ to 
$q=p_{\mu}-p_{\pi}$ that is conjugate to $\delta x$. We have then     
\begin{align}
I(p_{\pi},\delta x)&= {2 \over \pi} \int d^4 q  \, \theta(q^0+p_{\pi}^0) ((p_{\pi}+q)\!\cdot\! p_{\nu})
\text {Im}\left[1 \over
 (q+p_{\pi})^2-m_{\mu}^2-i\epsilon\right] e^{iq \cdot \delta x }\nonumber \\
&=(p_{\pi}\!\cdot\! p_{\nu}){2 \over \pi}\int d^4 q \,  \theta(q^0+p_{\pi}^0) 
\text {Im}\left[1 \over
 (q+p_{\pi})^2-m_{\mu}^2-i\epsilon\right] e^{iq \cdot \delta x }\nonumber \\
&+{2 \over \pi} \int d^4 q \,  \theta(q^0+p_{\pi}^0) (q \!\cdot\! p_{\nu})
\text {Im}\left[1 \over
 (q+p_{\pi})^2-m_{\mu}^2-i\epsilon\right] e^{iq\! \cdot\! \delta x }, 
\end{align}
and we separate the integration region into two parts:  
\begin{align}
&I(p_{\pi},\delta x)=I_1(p_{\pi},\delta x)+I_2(p_{\pi},\delta x), \nonumber\\
&I_1(p_{\pi},\delta x)=\left\{p_{\pi}\! \cdot\! p_{\nu}+p_{\nu}\!\cdot\! \left(-i{\partial \over
 \partial \delta x}\right)\right\} \tilde I_1,\nonumber\\
&\tilde I_1={2 \over \pi}\int d^4 q \,  \theta(q^0)\text {Im}\left[1 \over
 (q+p_{\pi})^2-m_{\mu}^2-i\epsilon\right] e^{iq \cdot \delta x }, \nonumber  \\
&I_2(p_{\pi},\delta x)= {2 \over \pi} \int_{-p_{\pi}^0}^{0}d q^0 d^3 q\, p_{\nu}\!\cdot\! (p_{\pi}+q)\text {Im} \left[1 \over
 (q+p_{\pi})^2-m_{\mu}^2-i\epsilon\right] e^{iq \cdot \delta x }.
\end{align}
$I_1(p_{\pi},\delta x)$ is the integral of the region $0 \leq q^0$ and 
has a singularity and
$I_2(p_{\pi},\delta x)$ is the integral of the region $-p_{\pi}^0 \leq
q^0 \leq 0$ and is regular. Although the large finite-distance  correction is 
derived from the singular term $I_1(p_{\pi},\delta x)$, $I_1$ does not 
contribute to the total probability at an infinite time  for the plane
waves. Conversely  $I_2(p_{\pi},\delta x)$ contributes to the total
probability without finite-distance correction for the plane
waves. Consequently by expressing  $I$ into the sum of $I_1$ and $I_2$, the
correction at finite distance is found easily.   
So the physical quantity at the
finite distance that reflects interference is computed using the most singular
term of $I_1$.   

\subsubsection{Correlation function }
The denominator of the integrand of $\tilde I_1$ is expanded in
$p_{\pi}\!\cdot\! q$
and  the $\tilde I_1$ is written in the form 
\begin{align}
&\tilde I_1(p_{\pi},\delta x)={2 \over \pi}\int d^4 q \, \theta(q^0)~ \text {Im}\left[{1 \over
q^2+m_{\pi}^2-m_{\mu}^2+2q\!\cdot\! p_{\pi}-i\epsilon}\right] e^{iq \cdot \delta x } \nonumber\\
&={2 \over \pi}\int d^4 q\,  \theta(q^0)~ \text {Im}\left[{1 \over
 q^2+{\tilde m}^2-i\epsilon} -2p_{\pi} \!\cdot\! q\left({1 \over
 q^2+{\tilde m}^2-i\epsilon}\right)^2 +\cdots \right]e^{iq \cdot\delta x } \nonumber \\
&={2 \over \pi}\int d^4 q  \,\theta(q^0)\left\{1-2p_{\pi}\!\cdot\! \left(-i{\partial \over \partial \delta
 x}\right) {\partial  \over \partial {\tilde m}^2}+\cdots \right\}\,\text {Im}\left[ {1 \over
 q^2+{\tilde m}^2-i{\epsilon}} \right]e^{iq \cdot\delta x } \nonumber\\
&=2  \left\{1 -2p_{\pi} \!\cdot\!\left(-i{\partial \over \partial \delta
			  x}\right) {\partial  \over \partial {\tilde m}^2}+\cdots \right\}
\int d^4 q \, \theta(q^0)\delta (q^2+{\tilde m}^2) e^{iq \cdot\delta x },\label{singular-function}
\end{align}
where 
\begin{eqnarray}
{\tilde m}^2=m_{\pi}^2-m_{\mu}^2.
\end{eqnarray}
The expansion in $2q\!\cdot\! p$ of 
Eq.\,$(\ref{singular-function})$  converges  in the region
\begin{eqnarray}
{2p_{\pi}\!\cdot\! q \over q^2+{\tilde m}^2} < 1.
\end{eqnarray}
Here $q$ is the integration variable and varies. So we  evaluate  
the series after the integral and find the condition for the convergence. 
 We  find later that the series after the momentum integration 
converges in the region
\begin{eqnarray}
{2p_{\pi}\!\cdot\! p_{\nu} \over {\tilde m}^2} \leq 1.
\label{convergence-condition}
\end{eqnarray}
So the following result that is obtained using  this expansion is
applied in this kinematical region. In the outside of
this region, the evaluation of the integrals $I_1$ and $I_2$ separately 
is not useful and $I$ is integrated directly. 

The formula for a relativistic field of the imaginary mass   
\begin{align}
\int d^4 &q  \, \theta(q^0) \delta(q^2+{\tilde m}^2)e^{iq \cdot\delta x }
= (2\pi)^3i\left[{1
 \over 4\pi}\delta(\lambda)\epsilon(\delta t) +f_{short}\right],\nonumber \\
f_{short}&=-{i {\tilde m} \over
 8\pi \sqrt{-\lambda}} \theta(-{\lambda})\left\{N_1\left(\tilde m \sqrt{
 -\lambda}\right)-i\epsilon(\delta t) J_1\left(\tilde m \sqrt{ -\lambda}\right)\right\} \nonumber \\
&-\theta(\lambda){i
 \tilde m \over
 4\pi^2\sqrt{\lambda}}K_1\left(\tilde m\sqrt{\lambda}\right),\label{singular-function-f}
\end{align}
where $N_1$, $J_1$, and $K_1$ are Bessel functions, is substituted to Eq.\,$(\ref{singular-function})$.
\begin{figure}[t]
\centering{
 \includegraphics[scale=.5]{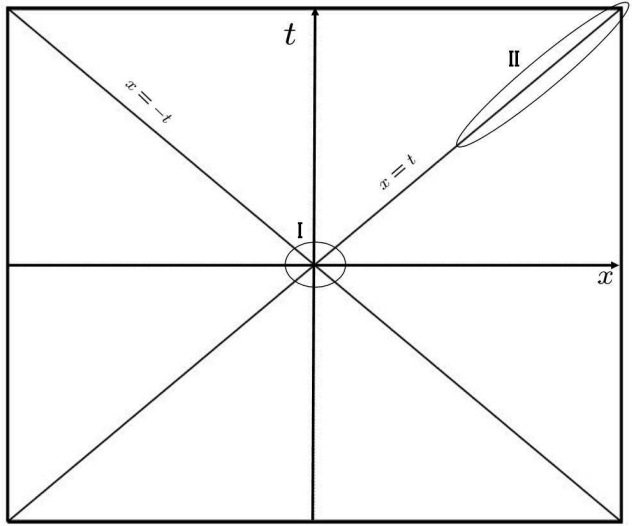}
}
\caption{The region I corresponds to short-range correlation where $t
 \sim x \sim 0$. On the other hand the region II corresponds to
 long-range correlation where $t^2 - x^2 \sim 0$ and both $t$
 and $x$ can be macroscopic.}
\label{fig:light-cone}
\end{figure}
The first term in the right hand side of
Eq.\,$(\ref{singular-function-f})$ is the most singular term and the
second and third terms have  singularity  of the form $1/\lambda$ around
$\lambda =0$ and decrease as $e^{-\tilde m\sqrt{|\lambda}|}$ or
oscillates as  $e^{i\tilde m \sqrt{|\lambda}|}$. The singular functions
and regular functions
behave differently and are expressed in Fig.\,\ref{fig:light-cone} for 
one space 
dimension.  The singular functions have the value around  the light
cone and the regular functions have finite value in  small area around
the origin. Since the light cone  is extended  in macroscopic area, the
light-cone singularity leads the correlation function to become
long-range. The long-range correlation function from the light-cone
singularity and the short-range correlation  function from the regular
function are computed next.
Thus the correlation function $\tilde I_1$ becomes 
long-range only along the light-cone region and decreases exponentially
or oscillates rapidly in other directions. So
     $\tilde I_1(p_{\pi},\delta
x)$ then is written in the form 

\begin{align}
\tilde I_1(p_{\pi},\delta x)
=  2(2\pi)^3i\left\{1 -2p_{\pi} \!\cdot\!\left(-i{\partial \over \partial
 \delta x}\right) {\partial  \over
 \partial {\tilde m}^2}+\cdots \right\}\left( {1 \over 4\pi}\delta(\lambda)\epsilon(\delta
 t)+f_{short}\right).
\end{align}

Next  $I_2$ is evaluated. For $I_2$ we use the momentum $\tilde
q=q+p_{\pi}$ and to write in the form
\begin{align}
& I_2(p_{\pi},\delta x)=\frac{2}{\pi} \int_{0< \tilde
 q^0<p_{\pi}^0} d^4 \tilde q \, (p_{\nu} \!\cdot\!\tilde q)  \text {Im}\left[\frac{1}
 {\tilde q^2-m_{\mu}^2-i\epsilon}\right] e^{i(\tilde q-p_{\pi}) \cdot\delta x } 
\nonumber\\
&=e^{i(-p_{\pi}) \cdot\delta x }\left\{p_{\nu} \!\cdot\!\left(-i{\partial \over \partial
 \delta x}\right)\right\}  {2 \over \pi}\int_{0< \tilde q^0
 <p_{\pi}^0 } d^4 \tilde{q}\,  \pi \delta( q^2-m_{\mu}^2)  e^{i\tilde q \cdot\delta x } 
\nonumber \\
&= e^{-i p_{\pi} \cdot\delta x}\left\{p_{\nu} \!\cdot\!\left(-i{\partial \over \partial
 \delta x}\right)\right\} \int{ d^3q \over
  \sqrt{q^2+m_{\mu}^2}} \theta\left(p_{\pi}^0-\sqrt{q^2+m_{\mu}^2} \right)
 e^{iq \cdot\delta x }.\label{normal-term}
\end{align}
The regular part $I_2$ has no singularity because the integration domain is
finite and becomes short-range.  

Thus the first term in $\tilde I_1$ gives a finite long-distance 
correlation  and the rests, the second
term in $I_1$ and $I_2$, give short distance correlations. 
The correlation function, $\tilde \Delta_{\pi,\mu}(\delta t,\delta {\vec x})$ 
has a singular term and a regular term and  is written in the form 
\begin{align}
&\tilde \Delta_{\pi,\mu}(\delta t,\delta {\vec x})={1 \over (2\pi)^3} {1 \over E(p_{\pi})}\Biggl[\left\{p_{\pi}\! \cdot\! p_{\nu}+p_{\nu}\!\cdot\!\left(-i\frac{\partial}{
 \partial \delta x}\right)\right\}2(2\pi)^3i\nonumber\\ 
& \left\{1 -2p_{\pi} \!\cdot\! \left(-i{\partial \over \partial
 \delta x}\right) {\partial  \over
 \partial {\tilde m}^2}+\cdots \right\} \left( \frac{1}{4\pi}\delta(\lambda)  
\epsilon(\delta t)+f_{short}\right) + I_2 \Biggr],\label{muon-correlation-total}
\end{align}
where the dots stand for the higher order terms.

\subsection{Integration of spatial coordinates   }
Next, the coordinates ${\vec x}_1$ and ${\vec x}_2$ are integrated in
\begin{align}
&\int d{\vec x}_1 d{\vec x}_2e^{i\phi(\delta x)}e^{-\frac{1}{2\sigma_{\nu} }\left({\vec x}_1-{\vec X}_{\nu} -{\vec
v}_\nu(t_1-T_{\nu})\right)^2}e^{-{1 \over 2\sigma_{\nu} }\left({\vec x}_2-{\vec
X}_{\nu} -{\vec v}_\nu(t_2-T_{\nu})\right)^2}\nonumber\\
&\times \tilde \Delta_{\pi,\mu}(\delta t,\delta{\vec x}).\label{lightcone-integration1}
\end{align}
The derivative $i{\partial \over \partial x}$ in the above integral
is  computed using the integration by part  as
\begin{align}
\int d x \, e^{i(\phi(\delta x) -p_{\pi}\cdot\delta x
 )}i\left(-{\partial \over \partial x}f(x)\right)&=\int dx \, i\left({\partial \over \partial x}
e^{i(\phi(\delta x) -p_{\pi} \cdot\delta x)}\right)f(x)\nonumber\\ 
&= \int dx (p_{\pi}-k_{\nu})e^{i(\phi(\delta x)-p_{\pi}\cdot\delta x)}f(x),
\end{align}
where a function $f(x)$ is an arbitrary function and 
Eq.\,(\ref{derivative-phase}) was used.
\subsubsection{Singular terms: long-range correlation}
The most singular term in Eq.~$(\ref{lightcone-integration1})$ is   
\begin{eqnarray}
\label{singular-correlation}
J_{\delta(\lambda)}&=&\int d{\vec x}_1 d{\vec x}_2e^{i\phi(\delta x)}e^{-{1 \over 2\sigma_{\nu} }\left({\vec x}_1-{\vec X}_{\nu} -{\vec
v}_\nu(t_1-T_{\nu})\right)^2}e^{-{1 \over 2\sigma_{\nu} }\left({\vec x}_2-{\vec
X}_{\nu} -{\vec v}_\nu(t_2-T_{\nu})\right)^2} \nonumber \\
&\times &{1 \over
4 \pi}\delta(\lambda)\epsilon(\delta t)  
\end{eqnarray}
and is rewritten using the center coordinate $X^\mu=\frac{
x_1^\mu+x_2^\mu}{2}$ and the relative coordinate
$\vec{r}=\vec{x}_1-\vec{x}_2$ in the form,
\begin{eqnarray}
\label{singular-correlation-center}
J_{\delta(\lambda)}&=&\int d{\vec X} d{\vec r}\, e^{i\phi(\delta x)}e^{-\frac{1}{\sigma_{\nu} }\left({\vec X}-{\vec X}_{\nu} -{\vec
v}_\nu(X^0-T_{\nu})\right)^2}e^{-\frac{1}{4\sigma_{\nu} }\left({\vec r}
-{\vec v}_\nu\delta t\right)^2} \nonumber \\
&\times &\frac{1}
{4 \pi}\delta(\lambda)\epsilon(\delta t).  
\end{eqnarray}
The center coordinate ${\vec X}$ is integrated  easily and $J_{\delta(\lambda)}$ becomes
the integral of the   transverse and longitudinal component $({\vec
r}_T,r_l)$ of the relative coordinates,   
\begin{align}
\epsilon(\delta t) (\sigma_{\nu}\pi)^{\frac{3}{2}} \int d{\vec r}_Td r_l \, e^{i\phi(\delta t,{\vec
 r})-\frac{1}{4\sigma_{\nu} }({{\vec r}_T}^{\,2}  +(r_l-{
 v}_{\nu}\delta t)^2)}\frac{1}{4\pi}\delta (\delta t^2-{{\vec r}_T}^{\,2} -{{\vec r}_l}^{\,2}).
\label{lightcone-integration-s}
\end{align}
Finally this  is computed in the form 
\begin{eqnarray}
J_{\delta(\lambda)}&=&{(\sigma_{\nu}\pi)}^{\frac{3}{2}}
 \frac{\sigma_{\nu}}{2}{1 \over |\delta t|
 }\epsilon(\delta t)e^{i\bar \phi_c(\delta t)-\frac{m_{\nu}^4}{
 16\sigma_{\nu} E_{\nu}^4} {\delta t}^2}\nonumber\\
 &\approx& {(\sigma_{\nu}\pi)}^{\frac{3}{2}} \frac{\sigma_{\nu}}{2}
  \frac{1}{|\delta t|
  }\epsilon(\delta t)e^{i\bar \phi_c(\delta t)}\label{lightcone-integration2-2}. 
\end{eqnarray}

The next term in Eq.\,$(\ref{lightcone-integration1})$ is from  
${1/\lambda}$. 
We have     
\begin{eqnarray}
J_{1/\lambda}&=&\int d{\vec x}_1 d{\vec x}_2e^{i\phi(\delta x)}e^{-{1 \over 2\sigma_{\nu} }\left({\vec x}_1-{\vec X}_{\nu} -{\vec
v}_\nu(t_1-T_{\nu})\right)^2}e^{-\frac{1}{2\sigma_{\nu} }\left({\vec x}_2-{\vec
X}_{\nu} -{\vec v}_\nu(t_2-T_{\nu})\right)^2} \nonumber \\
&\times &{i  \over 4\pi^2 \lambda},
\label{lightcone-integration4}
\end{eqnarray}
which becomes 
\begin{eqnarray} 
J_{1/\lambda}&\approx& {(\sigma_{\nu}\pi)}^{\frac{3}{2}} \frac{\sigma_{\nu}}{2} \left(\frac{1}{
 \pi \sigma_{\nu} p_{\nu}^2}\right)^{\frac{1}{2}} e^{-\sigma_{\nu}p_{\nu}^2}
\frac{ 1}{|\delta t| }e^{i\bar
\phi_c(\delta t)}.
\label{lightcone-integration4-2} 
\end{eqnarray} 
This term has the universal $|\delta t|$ dependence but its magnitude is much
smaller than that of $J_{\delta(\lambda)}$ and is negligible in the present decay mode.  

From Eqs.\,$(\ref{lightcone-integration2-2})$
and $(\ref{lightcone-integration4-2})$, the singular terms
$J_{\delta(\lambda)}$ and $J_{1/\lambda}$ have the slow  phase
$\bar \phi_c(\delta t)$ and the magnitudes that are inversely proportional
to the time difference. Thus these terms are long-range with the small
angular velocity and are
insensitive to the ${\tilde m}^2$.  These properties of the
time dependent correlation functions $J_{\delta(\lambda)}$ are most
important for the neutrino probability to have the universal
long-distance
 behavior and also hold
for  the general wave packets of
 Eq.\,$(\ref{neutrino-position-gamplitudeG})$ which satisfy
 Eq.\,$(\ref{real-even-wavefunction} )$.

{\bf Theorem}

The singular part $J_{\delta(\lambda)}$ of the correlation function has
the slow  phase that is determined by the absolute value of the neutrino
mass
and the 
magnitude that is  inversely proportional
to the time difference at the large distance, of the form 
Eq.\,$(\ref{lightcone-integration2-2})$. The phase is given by the sum
of $\bar \phi_c(\delta t)$ and the small correction, which is
inversely proportional to the neutrino energy in general systems 
and becomes $1/E^2$ if the neutrino wave
 packet is invariant under the time inversion and is real. 

{\bf (Proof: General cases including spreading of wave packet
)}

We prove the theorem for 
$J_{\delta(\lambda)}$ which is written in the form,
\begin{eqnarray}
\label{singular-correlation-centerG}
J_{\delta(\lambda)}=\int  d{\vec r}\, e^{i\phi(\delta x)}
\tilde w( \left({\vec r} -{\vec
v}_\nu\delta t\right)) \times \frac{1}
{4 \pi}\delta(\lambda)\epsilon(\delta t),  
\end{eqnarray}
where $\tilde w({\vec x}-{\vec v}t)$ is the wave packet in the
coordinate representation.
Since the light-cone singularity is extended in large time, the spreading
effect becomes important and is included in 
the time-dependent correlation function 
$J_{\delta(_\lambda)}$.  The wave function $w({\vec x}-{\vec v}t)$ with
the spreading effect is expressed in the following form 
\begin{eqnarray}
& &w({\vec x}-{\vec v}t)=\int dk_l d{\vec k}_T \,e^{ik_l(x_l-v_{\nu}t)+i{\vec k}_T\cdot{\vec
 x}_T+iC_{ij}k_T^ik_T^jt} w(k_l,{\vec k}_T),\\
& &C_{ij}=C_0 \delta_{ij},~C_0={1 \over 2E},
\end{eqnarray}
instead of Eq.$(\ref{fourier-transform1})$.
The  quadratic term of ${\vec
k}$ in the expansion of $E\left({\vec p}+{\vec k}\right)$ is included
and this makes the wave packet spread with time. The
coefficient $C_{ij}$ in the longitudinal direction is negligible for 
the neutrino and is neglected.  
$\tilde w(r_l-v_{\nu}\delta t,{\vec r}_T)$ is written   with this
$w({\vec x}-{\vec v}t)$ as
\begin{align}
\tilde w(r_l-v_{\nu}\delta t,{\vec r}_T)&=\int d{\vec X} w\left({\vec X}+\frac{\vec r}{2}\right)w^{*}\left({\vec X}-\frac{\vec r}{2}\right) \nonumber\\
&=\int dk_l d{\vec k}_T e^{ik_l(r_l-v_{\nu}\delta t)+i{\vec k}_T\cdot{\vec
 r}_T+ic_0({\vec k}_T^2)\delta t} |w(k_l,{\vec k}_T)|^2.
\end{align}
Then the correlation function becomes 
 \begin{align}
\label{singular-correlation-centerG4}
J_{\delta(\lambda)}&=\int dr_l d{\vec r}_T e^{i\phi(\delta t,r_l)}
\tilde w( r_l -v_\nu\delta t,{\vec r}_T){1 \over
4 \pi} \left\{1+\sum_{n=1} {1 \over n!}(-{{\vec r}_T}^{\,2})^n \left(\frac{\partial}
{\partial (\delta t )^2}\right)^n\right\}\nonumber\\
&\  \times \delta(\delta t^2-r_l^2) 
\epsilon(\delta t) \nonumber\\
&=  \int dr_l  d{\vec r}_T dk_l d{\vec k}_T e^{i\phi(\delta
  t,r_l)+ik_l(r_l-v_{\nu} \delta t)+i{\vec k}_T\cdot{\vec r}_T+iC_0{\vec k}_T^2 \delta
  t}|w(k_l,{\vec k}_T)|^2 \nonumber \\
&\ \times {1 \over
4 \pi} \left\{1+\sum_{n=1} {1 \over n!}(-{{\vec r}_T}^{\,2})^n \left(\frac{\partial}
{\partial (\delta t )^2}\right)^n\right\}  \delta(\delta t^2-r_l^2) 
\epsilon(\delta t)
\nonumber \\
&=  \int dr_l dk_l  e^{i\phi(\delta
  t,r_l)+ik_l(r_l-v_{\nu} \delta t)} d{\vec r}_T  d{\vec k}_T
e^{+i{C_0{\vec k}_T^2 \delta
  t}}|w(k_l,{\vec k}_T)|^2 \nonumber \\
&\ \times{1 \over 4 \pi} \left\{1+\sum_{n=1} {1 \over n!}({\partial^2 \over
		    (\partial {\vec k}_T)^{\,2}})^n \left(\frac{\partial}
{\partial (\delta t )^2}\right)^n\right\} e^{i{\vec k}_T\cdot{\vec r}_T}  \delta(\delta t^2-r_l^2) 
\epsilon(\delta t).
\end{align}
The variables ${\vec r}_T$ are integrated first and ${\vec k}_T$ are integrated
next. Then we have the expression 
\begin{align}
J_{\delta(\lambda)}&=  \int dr_l dk_l  e^{i{\phi}(\delta
  t,r_l)+ik_l(r_l-v_{\nu} \delta t)}  
|w(k_l,0)|^2  \nonumber \\
& \times {1 \over 4 \pi} \left\{1+\sum_{n=1} {1 \over n!}(-2iC_0 \delta t)^n \left(\frac{\partial}
 {2 \delta t \partial \delta t }\right)^n\right\} (2\pi)^2 \delta(\delta t^2-r_l^2) 
\epsilon(\delta t).
\end{align}
Using the following identity 
\begin{eqnarray}
(2\delta t)^n\left({\partial \over 2\delta t \partial \delta t}\right)^n=\left({\partial
 \over \partial \delta t}\right)^n+O\left({1 \over \delta t}\right) \left(\frac{\partial}{\partial \delta t}\right)^{n-1}
\end{eqnarray}
and taking the leading term in ${1/\delta t}$, we have the final
expression of the correlation function at the long-distance region
\begin{eqnarray}
& &
\label{singular-correlation-centerF}
J_{\delta(\lambda)}= \pi e^{-C_0 p } \epsilon(\delta t){e^{i\bar{\phi}_c(\delta
  t)} \over 2\delta t} \int  dk_l  e^{k_l(i(1-v_{\nu}) \delta t +C_0)}  
|w(k_l,0)|^2.   
\end{eqnarray}

Hence $J_{\delta (\lambda)}$ in Eq.\,$(\ref{singular-correlation-centerF}  )$
becomes  almost the same form as 
Eq.\,$(\ref{lightcone-integration2-2})$ and the slow phase
$\bar \phi_c(\delta t)$ is modified slightly and the magnitude that is 
inversely proportional
to the time difference. $J_{\delta(\lambda)}$ has the universal form for
the general wave packets. By expanding the exponential factor and taking
the quadratic term of the exponent, the above   integral is written in 
the form 
\begin{align}
\label{singular-correlation-centerFc}
& \int  dk_l  (1 +{k_l(i(1-v_{\nu}) \delta
  t +C_0)}+{1 \over 2!}{(k_l(i(1-v_{\nu}) \delta
  t +C_0) )^2})  
|w(k_l,0)|^2   \nonumber \\
&~=w_0(1+d_1(i(1-v_{\nu}) \delta
  t +C_0)+{d_2 \over 2!}(i(1-v_{\nu}) \delta
  t +C_0)^2)  \nonumber \\
&~=w_0\left(1+C_0d_1+{d_2 \over
 2!}C_0^2-(1-v_{\nu})^2{\delta t}^2\right)+i(d_1(1-v_{\nu})\delta t
 +d_2C_0(1-v_{\nu}) \delta t ).
\end{align}
We substitute this expression into the correlation function, and  we have    
\begin{align}
J_{\delta(\lambda)}=\pi e^{-C_0 p }\omega_0(1+\gamma) \epsilon(\delta t){e^{i\bar{\phi}_c(\delta
  t)(1+\delta)} \over 2\delta t}
, \ w_0= \int d k_l |w(k_l,0)|^2,
\end{align}
where the correction terms are given by
\begin{eqnarray}
& &\delta={d_1 \over E}+\frac{d_2}{2}{1 \over E^2},~
\gamma={d_1 \over 2E}+{d_2\over
 2!}\left({1 \over 2E}\right)^2-(1-v_{\nu})^2{\delta t}^2, \nonumber \\
& &d_1= \frac{1}{w_0}\int d k_l k_l |w(k_l,0)|^2,~d_2= \frac{1}{w_0}\int d k_l
 k_l^2|w(k_l,0)|^2.
\end{eqnarray}
In the wave packets of time reversal invariance, $|w(k_l,0)|^2$ is the
even function of $k_l$. Hence $d_1$ vanishes 
\begin{eqnarray}
d_1=0,
\end{eqnarray}
 and the
correction are 
\begin{eqnarray}
\delta=\frac{d_2}{2}{1 \over E^2},\ 
\gamma={d_2 \over 2!}\left({1 \over 2E}\right)^2-(1-v_{\nu})^2{\delta t}^2.
\end{eqnarray}
{\bf Q.E.D.}

The light-cone singularity is along ${\delta t}^2-{\delta {\vec x}}^2=0$,  
which is so close to the neutrino path that it gives a
finite   contribution to the integral
Eq.$(\ref{lightcone-integration1})$.  Since the light-cone singularity is
real, the integral is sensitive  only to the neutrino phase and shows  
interference of the neutrino. 
\subsubsection{Regular terms: short-range correlation}
Next we study the regular terms. The regular term is short-range and the
spreading effect is ignored and the Gaussian wave packet is
studied. First term is  $f_{short }$ in
$I_1$ and is expressed by Bessel functions. We  have   
\begin{eqnarray}
L_1&=&\int d{\vec x}_1 d{\vec x}_2\, e^{i\phi(\delta x)}e^{-\frac{1}{2\sigma_{\nu} }\left({\vec x}_1-{\vec X}_{\nu} -{\vec
v}_\nu(t_1-T_{\nu})\right)^2-{1 \over 2\sigma_{\nu} }\left({\vec x}_2-{\vec
X}_{\nu} -{\vec v}_\nu(t_2-T_{\nu})\right)^2}  \nonumber \\
& & \times
f_{short}. 
\label{lightcone-integration2-1}
\end{eqnarray}
$L_1$ is evaluated at a large $|\delta t|$ and we have 
\begin{align}
L_1 = ({\pi \sigma_{\nu}})^{\frac{3}{2}}e^{iE_{\nu}\delta t}
\int d{\vec r} \, e^{-i(\vec p_{\nu} \cdot \vec r)-{1 \over 4\sigma_{\nu} }({\vec r} 
- {\vec v}_\nu\delta t)^2}  f_{short},~{\vec r}={\vec x}_1-{\vec x}_2.
\end{align}
Here the integration is made in the space-like region $\lambda <0$. It is 
convenient to write
\begin{eqnarray}
r_l=v_{\nu}\delta t +\tilde r_l,
\end{eqnarray}
and  to write $\lambda$ in the form 
\begin{align}
\lambda =\delta t^2-{\vec r}_l^{\,2}-{\vec r_T^{\,2}} =\delta
 t^2-(v_{\nu}\delta t + \tilde r_l)^{2}-{\vec r_T^{\,2}}
\approx -2 v_{\nu}\tilde r_l \delta t -\tilde r_l^2-{\vec
 r_T^{\,2}}.
\end{align}
The $L_1$ for the large $|\delta t|$ in the space-like region is written
with the asymptotic expression of the Bessel function and becomes 
\begin{eqnarray}
L_1&=&({\pi
 \sigma_{\nu}})^{\frac{3}{2}}e^{i(E_{\nu}-p_{\nu}v_{\nu})\delta t}
\int d{\vec r}_T d \tilde r_l \, e^{-i(  p_{\nu} \tilde r_l)-\frac{1}{4\sigma_{\nu} }(
\tilde r_l^2+ {\vec r}_T^{\,2})} {i\tilde m \over 4\pi^2}\left({\pi \over
						       2\tilde m}\right)^{\frac{1}{2}}
\nonumber \\  
& & \times\left({ 1  \over {+2 v_{\nu}\tilde r_l \delta t +\tilde r_l^2+
{\vec r_T^{\,2}} }}\right)^{\frac{3}{4}}
e^{i\tilde m  \sqrt{2 v_{\nu}\tilde r_l |\delta t| +\tilde r_l^2+{\vec r_T^{\,2}}
}} . 
\end{eqnarray}
By the Gaussian integration around ${\vec r}_t={\vec
0}$, $\tilde r_l=-i2\sigma_{\nu}p_{\nu}$, we have the asymptotic expression of
$L_1$ at a large $|\delta t|$
\begin{align}
L_1&=({\pi \sigma_{\nu}})^{\frac{3}{2}}\tilde L_1,\nonumber\\
\tilde L_1&=
e^{i(  E_{\nu}-p_{\nu}
 v_{\nu})\delta t}  e^{-  \sigma_\nu p_{\nu}^2}{i \tilde m \over 4\pi^2}\left({\pi
 \over 2 \tilde m}\right)^{\frac{1}{2}}
 \left({ 1 \over {+2 v_{\nu}2 \sigma_{\nu} p_{\nu}
\delta t }}\right)^{{\frac{3}{4}}}
e^{i\tilde m \sqrt{2 v_{\nu}\sigma_{\nu} p_{\nu}|\delta t|}}.\label{asymptotic-expansionL_1}
\end{align}
Obviously $L_1$ oscillates  fast as $e^{i\tilde mc_1|\delta t|^{\frac{1}{2}}}$ where $c_1$
 is determined by $p_{\nu}$ and $\sigma_{\nu}$ and is short-range.
 The integration carried out with a different  stationary value of $r_l$ 
which takes into account the last term in the right-hand side gives  almost
 equivalent result. The integration of $L_1$ in the time-like region 
$\lambda >0$
 is carried in a similar manner and $L_1$ decreases  with time as
 $e^{-\tilde mc_1|\delta t|^{\frac{1}{2}}}$ and 
final  result after the time integration is almost the same as that of 
the space-like region.
 It is noted that the long-range term which appeared 
from the isolated ${1/\lambda}$ singularity in 
Eq.\,$(\ref{lightcone-integration4-2}) $ does  not exist  in $L_1$ in
 fact. The reason for its absence is that the Bessel function decreases 
much faster in the  space-like region than ${1/\lambda}$ and
 oscillates much faster 
 than ${1/\lambda}$ in the time-like region. Hence the long-range
 correlation is not generated from the $L_1$  and 
 the light-cone singularity $\delta(\lambda) \epsilon(\delta t)$ and 
$1/{\lambda}$ are the
 only source of the long-distance correlation. 

Second term is from $I_2$, Eq.\,$(\ref{normal-term})$. We have this
 term, $L_2$, 
\begin{align}
L_2&=2 p_{\nu}\!\cdot\!(p_{\pi}-p_{\nu})(\pi\sigma_{\nu})^{\frac{3}
{2}}(4\pi\sigma_{\nu})^{\frac{3}{2}}{1 \over
  \left(2\pi\right)^3}\tilde L_2,\nonumber\\
\tilde L_2&= \int {d^3q \over 2\sqrt{q^2+m_{\mu}^2}} 
 e^{-i\left(E_\pi-E_{\nu}-\sqrt{q^2+m_{\mu}^2}-{\vec v}_{\nu}\cdot({\vec
 p}_{\pi}-{\vec q}-{\vec p}_{\nu})\right)\delta t}  \nonumber \\
&\times e^{ -{ \sigma_{\nu} ({\vec p}_{\pi}-{\vec q}-{\vec p}_{\nu})^2}} 
\theta \left(E_\pi-\sqrt{q_t^2+q_l^2+m_{\mu}^2} \right).\label{lightcone-integration5}
\end{align}

 The angular velocity of Eq.\,$(\ref{lightcone-integration5})$ in $L_2$
 varies with ${\vec q}$ and $L_2$ becomes to have a short-range
 correlation of the length, $2 \sqrt{
 \sigma_{\nu}}$, in the time direction.
So the $L_2$'s contribution to the total probability comes from the small
 $\delta t$ region and  this corresponds to the short-range component.

Thus the coordinate integration of $\tilde \Delta_{\pi,\mu}(\delta t,\delta\vec{x})$ is 
written in the form
\begin{align}
&\int d{\vec x}_1 d{\vec x}_2\,e^{i\phi(\delta x)}e^{-{1 \over 2\sigma_{\nu} }\left({\vec x}_1-{\vec X}_{\nu} -{\vec
v}_\nu(t_1-T_{\nu})\right)^2}e^{-\frac{1}{2\sigma_{\nu} }\left({\vec x}_2-{\vec
X}_{\nu} -{\vec v}_\nu(t_2 - T_{\nu})\right)^2} \tilde \Delta_{\pi,\mu}(\delta
t,\delta{\vec x})\nonumber \\
& =2i  {1 \over E_\pi} p_{\pi}\! \cdot\! p_{\nu}\left[ \left(1 +2p_{\pi}
 \!\cdot\! p_{\nu} \frac{\partial}{\partial {\tilde m}^2}+\cdots
 \right)e^{i\bar{\phi}(\delta t)}(J_{\delta(\lambda)}+L_1)+L_2\right]
\nonumber\\
&\approx 2i (\pi\sigma_{\nu})^{\frac{3}{2}}  {1 \over E_\pi} p_{\pi} \!\cdot\! p_{\nu} \left[\left(1
 +2p_{\pi} \!\cdot\! (-p_{\nu} ) 
{\partial  \over \partial {\tilde m}^2}+\cdots \right)\right. \nonumber \\ 
&\left.~~~~\times\left(\sigma_{\nu} {1 \over 2} e^{i\bar \phi_c(\delta t)} 
{\epsilon(\delta t) \over |\delta t|}
 +\tilde L_1\right) +\left( \frac{\sigma_\nu}{\pi}\right)^{\frac{3}{2}}(-i)\tilde L_2\right].\label{lightcone-integration1-1}
\end{align}

In the above equation, $p_{\nu}^2=m_{\nu}^2$ is neglected since this is 
extremely small compared to $\tilde{m}^2$, $p_{\pi}\!\cdot\! p_{\nu}$ and
$\sigma_{\nu}$. This is  neglected  also in most other places 
except the slow phase $\bar \phi(\delta t)$. The 
first term in the right-hand side of
Eq.\,$(\ref{lightcone-integration1-1})$ has the long-distance correlation and 
the second term has a short distance correlation. They  are separated
in a clear manner.

\subsubsection{Convergence condition}

At the end of this section, we study when our method is valid. In our
calculations the integration region is split into the one of finite
region and the other of infinite region and the separation of the
correlation function into the singular term and the regular 
term, which is equivalent to the separation of the correlation function
into the long-distance term and the short distance term,  is made next.
The long-distance term is generated from the light-cone
singularity, which is  obtained by the the expansion of
Eq.\,$(\ref{singular-function})$, so for our method to be valid, 
the convergence condition and the separability of the long-distance 
behavior should be satisfied. 

We study the convergence condition when the 
power series 
\begin{eqnarray} 
\label{power-series}
\sum_n (-2p_{\pi} \!\cdot\! p_{\nu} )^n {1 \over n!} 
\left({\partial  \over \partial {\tilde m}^2}\right)^n \tilde L_1,
\end{eqnarray}
becomes finite using the asymptotic expression of $\tilde
L_1$, Eq.\,$(\ref{asymptotic-expansionL_1})$, first.  The most
dangerous term
in $\tilde L_1$, is ${\tilde m}^{\frac{1}{2}}$. Other terms converge
when this converges. So we find  the
convergence condition from the series  
\begin{eqnarray}
S_1=\sum_n (-2p_{\pi} \!\cdot\! p_{\nu} )^n {1 \over n!} 
\left({\partial  \over \partial {\tilde m}^2}\right)^n ({\tilde m}^2)^{\frac{1}{4}}.
\end{eqnarray}
The  $S_1$ becomes into the form,
\begin{align}
S_1&=\sum_n \left({-2p_{\pi} \!\cdot\! p_{\nu} \over {\tilde m}^2 }\right)^n {1 \over
 n!} \left(n-\frac{1}{4}\right)!(-1)^n ({ \tilde m})^{\frac{1}{2}}  \nonumber\\
&\approx \sum_n \left(-{2p_{\pi} \!\cdot\! p_{\nu} \over {\tilde m}^2} \right)^n
 (-1)^n n^{-\frac{5}{4}}({ \tilde m})^{\frac{1}{2}} =\sum_n \left({2p_{\pi} \!\cdot\! p_{\nu}\over {\tilde m}^2} \right)^n n^{-\frac{5}{4}} 
({ \tilde
 m})^{\frac{1}{2}}.
\end{align} 
Hence the series converges in the kinematical region
Eq.\,$({\ref{convergence-condition}})$. At $2p_{\pi} \cdot p_{\nu}={\tilde
m}^2 $ $S_1$ becomes finite, and the value is expressed by the zeta function,
\begin{eqnarray}
S_1=\sum_n n^{-\frac{5}{4}} ({ \tilde m})^{\frac{1}{2}}=\zeta\left(\frac{5}{4}\right)  ({
 \tilde m})^{\frac{1}{2}}.
\end{eqnarray}
Hence in the region, Eq.\,$({\ref{convergence-condition}})$, the
correlation function has the singular terms and the total 
probability has  the long-range 
terms $J_{\delta(\lambda)}$ and $J_{1/\lambda}$.
In the outside of this region, the power series diverges and our method
does not work. $I$ is evaluated directly and has no long
range term. The $I$
obtained from the finite muon momentum is equivalent to the $I_2$.  

We study if the power series Eq.\,$(\ref{power-series})$ oscillates with
time $\sqrt{|\delta t|}$ rapidly next.
For this purpose we study
\begin{eqnarray}
S_2=\sum_n (-2p_{\pi} \!\cdot\! p_{\nu} )^n {1 \over n!} 
\left({\partial  \over \partial {\tilde m}^2}\right)^ne^{i\tilde m \sqrt{2
 v_{\nu}\sigma_{\nu} p_{\nu}|\delta t| }}.
\end{eqnarray}
When $S_2$ oscillates with the $\sqrt{|\delta t|}$, other terms in $\tilde
L_1$ oscillate with the $\sqrt{|\delta t|}$.
By the explicit calculations, we have 
 \begin{eqnarray}
S_2=
e^{i\tilde m\left|\sqrt{2
 v_{\nu}\sigma_{\nu} p_{\nu}|\delta t| }\right|\left(1-\frac{p_{\pi} \cdot p_{\nu}}
{{\tilde m}^2}\right)},
\end{eqnarray}
and the function oscillates with $\sqrt{|\delta t|}$ in the kinematical region
Eq.\,$({\ref{convergence-condition}})$. So the separating the light-cone
singular term from $I_1$ is valid and is used for the evaluation of the
finite-size correction of the probability at the finite distance. 


\section{Time-dependent probability}

Substituting Eqs.\,(\ref{light-phase}), (\ref{pi-mucorrelation}), and (\ref{lightcone-integration1-1}) into Eq.\,(\ref{probability-correlation}), we have 
the probability for measuring the neutrino at
a space-time position $({\vec X}_{\nu},T_{\nu})$, when the pion momentum distribution $\rho_{exp}({\vec
p}_{\pi})$ is known, in the following form  
\begin{align}
&\int d{\vec p}_{\pi} \rho_{exp}({\vec
p}_{\pi})  d{\vec p}_{\mu}  d{\vec p}_\nu  \sum_{s_1,s_2}|T|^2 
=g^2 m_{\mu}^2
 |N_{\pi\nu}|^2{(\sigma_{\nu}\pi)}^{\frac{3}{2}}{ 1 \over
 (2\pi)^6}{\sigma_{\nu}} \int {d^3
 p_{\pi} \over E_\pi}\rho_{exp} \label{total-probability2}({\vec p}_{\pi})\nonumber \\
&\times\int {d{\vec
 p}_\nu  \over E_{\nu}}   p_{\pi} \!\cdot\! p_{\nu} \int dt_1 dt_2
\left[    e^{i {m_{\nu}^2  \over 2E_{\nu}}\delta t} 
\frac{\epsilon(\delta t)}{|\delta t|}
 +{2 \tilde L_1 \over \sigma_{\nu}}+{2 \over \pi}\left( {\sigma_\nu \over \pi}\right)^{\frac{1}{2}}(-i)\tilde L_2\right]\nonumber\\
&\times  e^{-{1 \over 2 \sigma_{\pi} } \left({\vec X}_{\nu}
-{\vec X}_{\pi}+({\vec v}_{\nu}-\bar {\vec v}_{\pi})(t_1-T_{\nu}) + 
\bar{\vec{v}}_{\pi}(T_{\pi}- T_{\nu})\right)^2-{1 \over 2 \sigma_{\pi} }\left({\vec X}_{\nu}-{\vec X}_{\pi}+({\vec
v}_{\nu}-\bar{{\vec v}}_{\pi})(t_2-T_{\nu}) + \bar {\vec{v}}_{\pi}(T_{\pi}
- T_{\nu})\right)^2}.
\end{align}
From the pion coherence length obtained in the previous section, the
coherence condition, Eq.\,$(\ref{coherence-condition})$, is satisfied
and  the pion Gaussian parts are regarded as constant in $t_1$ and $t_2$,
 \begin{eqnarray}
& & e^{-\frac{1}{2\sigma_{\pi}}\left(\vec{X}_{\nu} - \vec{X}_{\pi} +
			     (\vec{v}_{\nu} - \bar {\vec{v}}_{\pi})(t_1 -
			     T_{\nu}) + \bar {\vec{v}}_{\pi}(T_{\pi} -
			     T_{\nu})\right)^2}\approx \text{constant in
}t_1,\nonumber   \\
& & e^{-\frac{1}{2\sigma_{\pi}}\left(\vec{X}_{\nu} - \vec{X}_{\pi} +
			     (\vec{v}_{\nu} - \bar {\vec{v}}_{\pi})(t_2 -
			     T_{\nu}) + \bar {\vec{v}}_{\pi}(T_{\pi} -
			     T_{\nu})\right)^2}\approx \text{constant in
}t_2,\label{coherence-conditions}
 \end{eqnarray}
when the integration on time $t_1$ and $t_2$ are made in a distance of 
our interest which is of order a few $100$ [m].  
The integration on $t_1,t_2$ will be made in the next section.

 When the above  conditions Eq.\,$(\ref{coherence-conditions})$ are
 satisfied, the area where the neutrino is produced is inside of the
 same pion and the neutrino waves are treated coherently and are capable
 of showing interference. In a much larger distance where this 
condition is not satisfied,  two positions can not be in the same pion
 and the interference disappears.

\subsection{Integrations on times}
Integration of the probability over the time $t_1$ and $t_2$ are carried
and probability at a finite T is obtained now. 
The time integral  of the slowly decreasing   term is  
\begin{eqnarray}
& &i  \int_0^{\text{T}} dt_1 dt_2  {e^{i {\omega_{\nu}}\delta t }
 \over |\delta t|}\epsilon(\delta t)   
= \text{T} g(\text{T},\omega_{\nu}),\ 
\omega_{\nu}={m_{\nu}^2 \over 2E_{\nu}},\label{probability1} 
\end{eqnarray}
where $g(\text{T},\omega_{\nu}) $ is 
\begin{align}
g(\text{T},\omega_{\nu})=-2\left(\text{Sin}~ x-{1 -\cos x \over
			    x}\right),\ x=\omega_{\nu} \text{T},\ \text{Sin}~x=\int_0^x dt {\sin~ t  \over t}.
\end{align}
The slope of $g(\text{T},\omega_{\nu})$ at $\text{T}=0$ is 
\begin{eqnarray}
{\partial \over \partial \text{T}}g(\text{T},\omega_{\nu})|_{\text{T}=0}
 ={\partial x \over \partial
 \text{T}}{\partial \over \partial x}\left.\left[-2\left(\text{Sin}~ x-{1 -\cos x \over
 x}\right)\right]\right|_{x=0}
= -\omega_{\nu}.
\end{eqnarray}
At the infinite time $\text{T}=\infty $, $g(\text{T},\omega_{\nu})$ becomes 
$g(\infty,\omega_{\nu})= {-\pi}$ 
that is cancelled with the short-range term of $I_1$ 
of  Eq.\,$(\ref{singular-function})$.
So it is convenient to subtract the asymptotic value from 
$g(\text{T},\omega_{\nu})$ and define $\tilde g(\text{T},\omega_\nu) $ 
\begin{eqnarray}
\tilde g(\text{T},\omega_\nu)= g(\text{T},\omega_\nu)-g(\infty,\omega_\nu).
\end{eqnarray}
We understand that the short-range part $L_1$ cancels with
$g(\infty,\omega_\nu) $ and write the total probability with 
$\tilde g(\text{T},\omega_\nu) $ and the short-range term from $ I_2$.

The  time integral of the short-range term, $ \tilde L_2$, is  
\begin{align}
&{2 \over \pi}\sqrt{\sigma_\nu \over \pi}\int dt_1 dt_2 \tilde
 L_2(\delta t) \nonumber\\
&= {2 \over \pi}\sqrt{\frac{\sigma_\nu}{\pi}} \int_0^{\text{T}} dt_1 dt_2 
 \int \frac{d^3q}{2\sqrt{q^2+m_{\mu}^2}} 
 e^{-i\left(E_{\pi}-E_{\nu}-\sqrt{q^2+m_{\mu}^2}-{\vec v}_{\nu}\cdot({\vec
 p}_{\pi}-{\vec q}-{\vec p}_{\nu})\right)\delta t}  \nonumber \\
&\ \ \ \times e^{ -{ \sigma_{\nu} ({\vec p}_{\pi}-{\vec q}-{\vec p}_{\nu})^2}}
 \theta\left(E_{\pi}-\sqrt{q_t^2+q_l^2+m_{\mu}^2}\right)
    \nonumber \\
&=\text{T} G_0,
\end{align}
where the constant $G_0$ is given in the integral   
\begin{eqnarray}
& &G_0=2 \sqrt{\sigma_\nu \over \pi}  \int {d^3q \over \sqrt{q^2+m_{\mu}^2}} \delta \left(E_{\pi}-E_{\nu}-\sqrt{q^2+m_{\mu}^2}-{\vec v}_{\nu}\cdot({\vec
 p}_{\pi}-{\vec q}-{\vec p}_{\nu})\right)  \nonumber
 \\
& &\times e^{ -{ \sigma_{\nu} ({\vec p}_{\pi}-{\vec q}-{\vec p}_{\nu})^2}}
 \theta\left(E_{\pi}-\sqrt{q_t^2+q_l^2+m_{\mu}^2}\right), 
\end{eqnarray}
and is estimated numerically. Due to the rapid oscillation in $\delta t$,
$ \tilde L_2$ contributes to the probability from the small $\delta t$
region and the integrations over the time becomes constant early with the
time interval T. This has no finite size  correction. The regular 
term $\tilde L_1$ is also the same.

\begin{figure}[t]%
\begin{center}
\includegraphics[angle=-90,scale=.5]{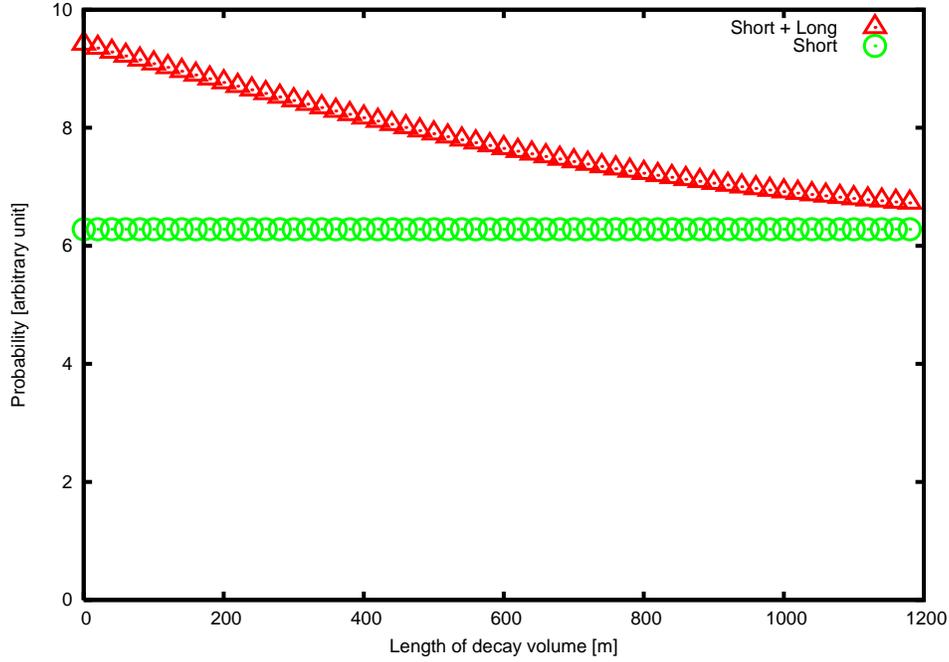}
   \end{center}
\caption{The  neutrino  probability in the forward direction   per time at a finite distance L is 
given. The constant shows the short-range normal term and   the long-range term is written on top of the normal term.  The horizontal axis 
shows the distance in~[m] and the probability  is
  of arbitrary unit. Clear excess is seen in the
 distance below 1200~[m]. The neutrino mass, pion energy, neutrino energy are
  1~[eV/$c^2$], 4~[GeV], and 800~[MeV].}
\label{fig:virtual-pi-singular}
\end{figure}%

\subsection{Total transition probability    }

Adding   the slowly decreasing part  and the short-range part,  we have 
the final expression of the total 
probability. The neutrino coordinate ${\vec X}_{\nu}$ 
of the final state  is integrated in Eq.($\ref{total-probability2})$ and a factor  $({\sigma_{\pi}\pi})^{\frac{3}{2}}$ emerges. 
This $\sigma_{\pi}$ dependence is cancelled  by 
the $({4\pi/\sigma_{\pi}})^{\frac{3}{2}}$ from  the
normalization Eq.\,$(\ref{probability})$ and the final result is 
independent from
$\sigma_{\pi}$.  The total transition 
probability is expressed in the form,  
\begin{eqnarray}
& &P=\text{T}g^2 m_{\mu}^2
 D_0  \sigma_{\nu}
\int {d^3
 p_{\pi} \over E_{\pi}}\rho_{exp}({\vec
p}_{\pi})\int {d^3 p_{\nu} \over E_{\nu}}  (p_{\pi}\! \cdot\! p_{\nu}) 
 [\tilde g(\text{T},\omega_{\nu}) 
 +G_0 ]
, \nonumber\\
& &
D_0=|N_{\pi\nu}|^2{(\sigma_{\nu}\pi)}^{\frac{3}{2}}
{(\sigma_{\pi}\pi)}^{\frac{3}{2}}{1 \over (2\pi)^6}={1 \over (2\pi)^3}
\label{probability-3}
\end{eqnarray}
where $\text{L} = c\text{T}$ is the length of decay region. 
Eq.\,$(\ref{probability-3})$
depends on the neutrino wave packet size $\sigma_{\nu}$.

 \begin{figure}[t]%
\begin{center}
  \includegraphics[angle=-90,scale=.5]{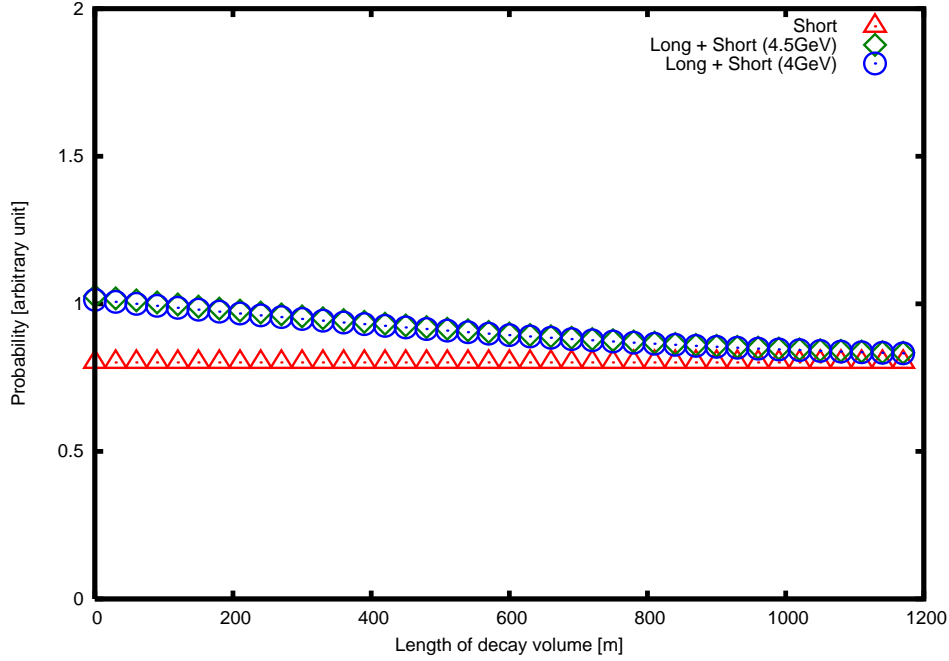}
   \end{center}
  \caption{The  total probability  integrated in the neutrino angle per
  time at a finite distance L is 
given. The constant shows the short-range normal term and   the long-range term is written on top of the normal term.  The horizontal axis 
shows the distance in~[m] and the probability  is
  of arbitrary unit. The excess becomes less clear than the forward
  direction, but is seen in the distance below 1200\,[m]. The neutrino mass, pion energy, neutrino energy are
  1.0~[eV/$c^2$], 4~[GeV] and 4.5~[GeV] , and 800~[MeV].}
 \label{fig:total-int-1}
\end{figure}%

\subsubsection{Neutrino angle integration}
In the normal term $G_0$ of Eq.\,($\ref{probability-3} $) the cosine of
neutrino angle $\theta$ is
determined approximately from 
\begin{eqnarray}
(p_{\pi}-p_{\nu})^2=p_{\mu}^2=m_{\mu}^2,
\end{eqnarray}
because the energy and momentum conservation is approximately satisfied 
in the normal term.  
Hence the product of the momenta is expressed by the masses
\begin{eqnarray}
p_{\pi}\!\cdot\! p_{\nu}={m_{\pi}^2-m_{\mu}^2 \over 2}, 
\label{on-shell-angle}
\end{eqnarray}
and  the cosine of the angle satisfies 
\begin{eqnarray}
1-\cos \theta= {m_{\pi}^2-m_{\mu}^2 \over 2|\vec{p}_{\pi}||\vec{p}_{\nu}|}-{m_{\pi}^2
 \over 2|\vec{p}_{\pi}|^2}.
\end{eqnarray}
The $\cos \theta $ is very close to 1.
On the other hand,  the long-range component of the neutrino probability, $\tilde g(\text{T},\omega_{\nu})$ of Eq.\,($\ref{probability-3} $), is derived from the light-cone singular
term. This term is present only
when the product of the momenta is in the convergence domain Eq.\,$({\ref{convergence-condition}})$.
Hence the long-range term is present in the kinematical region,  
\begin{eqnarray}
\label{long-kinematical}
|\vec{p}_{\nu}|(E_{\pi}-|\vec{p}_{\pi}|)\leq p_{\pi}\!\cdot\! p_{\nu} \leq {m_{\pi}^2-m_{\mu}^2 \over 2}. 
\end{eqnarray}

Since  the angular region of Eq.\,($\ref{long-kinematical}$) is 
slightly different from Eq.\,$(\ref{on-shell-angle})$ and it is impossible
to distinguish the latter  from the former region
experimentally, the neutrino angle is integrated.   
We integrate the neutrino angle of both terms separately.
We have  the normal term, $G_0$, in the form
\begin{align}
& \int \frac{d\vec{p}_\nu}{E_\nu} 
(p_\pi\cdot p_\nu) G_0  \nonumber\\
 &\simeq \int \frac{d\vec{p}_\nu}{E_\nu}
 (p_\pi\!\cdot\! p_\nu
 )2\sqrt{\frac{\sigma_\nu}{\pi}}\left(\frac{\pi}{\sigma_\nu}\right)^{\frac{3}{2}}\int \frac{d\vec{q}}{\sqrt{q^2 + m_\mu}} \nonumber \\
&\times \delta\left(E_\pi - E_\nu - \sqrt{q^2 + m_\mu^2}\right)\delta^{(3)}\left(\vec{p}_\pi - \vec{p}_\nu -
 \vec{q}\right)\theta\left(E_\pi - \sqrt{q^2 + m_\mu^2}\right) \nonumber\\
 &=
 \left(\frac{1}{\sigma_\nu}\right)(2\pi)^2\left({m_{\pi}^2-m_{\mu}^2 \over
 2}\right){1 \over |\vec{p}_{\pi}|}\int_{E_{\nu,min}}^{E_{\nu},max} dE_{\nu},
\end{align}
where 
\begin{eqnarray}
E_{\nu,min}={m_{\pi}^2-m_{\mu}^2 \over
 2(E_{\pi}+|\vec{p}_{\pi}|)},E_{\nu,max}={m_{\pi}^2-m_{\mu}^2 \over
 2(E_{\pi}-|\vec{p}_{\pi}|)}
\end{eqnarray}
and  the Gaussian function is approximated by the delta function for
the computational convenience. The angle is determined uniquely. We have  
for the long-range term, $\tilde{g}(\text{T},\omega_\nu)$, in the form
\begin{align}
 &\int \frac{d\vec{p}_\nu}{E_\nu} (p_\pi\!\cdot\! p_\nu)
 \tilde{g}(\text{T},\omega_\nu) \nonumber\\
&= 2\pi\int \frac{|\vec{p}_\nu|^2d|\vec{p}_\nu|}{E_\nu} \int_{\frac{E_\pi E_\nu - \frac{1}{2}(m_\pi^2 - m_\mu^2)}{|\vec{p}_\pi||\vec{
 p}_\nu|}}^{1}d\cos\theta(E_\pi E_\nu
 - |\vec{p}_\pi||\vec{ p}_\nu|\cos\theta)\tilde{g}(\text{T},\omega_\nu) \nonumber\\
&= 2\pi\int \frac{|\vec{p}_\nu|^2d|\vec{p}_\nu|}{E_\nu}\left[E_\pi E_\nu\cos\theta -
 \frac{1}{2}|\vec{p}_\pi|| \vec{p}_\nu| \cos^2\theta\right]_{\frac{E_\pi E_\nu - \frac{1}{2}(m_\pi^2 - m_\mu^2)}{|\vec{p}_\pi|
 |\vec{p}_\nu|}}^{1}\tilde{g}(\text{T},\omega_\nu) \nonumber\\
 & = 2\pi\int_{E_{\nu,min}}^{E_{\nu},max}  \frac{dE_\nu}{2|\vec{p}_\pi|}\left\{\frac{1}{4}\left(m_\pi^2 -
 m_\mu^2\right)^2 - (E_\pi E_\nu -
 |\vec{p}_\pi||\vec{p}_\nu|)^2\right\}\tilde{g}(\text{T},\omega_\nu),
\end{align}
where the angle is very close to the former value but is not uniquely
determined.
  Finally we have the energy dependent probability
\begin{eqnarray}
& &\frac{dP}{dE_\nu}=\text{T}g^2 m_{\mu}^2
 D_0  \int {d^3
 p_{\pi} \over E_{\pi}}\rho_{exp}({\vec
p}_{\pi}) 
 \frac{2\pi}{|\vec{p}_\pi|}\times\biggl[ {\pi}( m_\pi^2
 - m_\mu^2)  \nonumber\\
& &+{ \sigma_{\nu} \over 2}\left\{\frac{1}{4}\left(m_\pi^2 -
 m_\mu^2\right)^2 - (E_\pi E_\nu - |\vec{p}_\pi|
 |\vec{p}_\nu|)^2\right\}\tilde{g}(\text{T},\omega_\nu)\biggr]\label{total-probability-energy}.
\end{eqnarray}
\subsection{Comparison with measurements}
\subsubsection{Sharp pion momentum}
When initial pion has a discrete momentum ${\vec P}_{\pi}$, the
$\rho_{exp}({\vec p}_{\pi})$ is given as  
\begin{eqnarray}
\rho_{exp}({\vec p}_{\pi})=\delta({\vec p}_{\pi}-{\vec P}_{\pi})
\end{eqnarray}
and the energy-dependent probability for this pion is 
\begin{eqnarray}
& &\frac{dP}{dE_\nu}=\text{T}g^2 m_{\mu}^2
 D_0   {1 \over E_{\pi}} 
 \frac{2\pi}{|\vec{P}_\pi|}\biggl[ {\pi}( m_\pi^2
 - m_\mu^2) \label{total-probability-energy2}  \nonumber\\
& &+{ \sigma_{\nu} \over 2}\left\{\frac{1}{4}(m_\mu^2 -
 m_\pi^2)^2 - (E_\pi E_\nu - |\vec{P}_\pi|
 |\vec{p}_\nu|)^2\right\}\tilde{g}(\text{T},\omega_\nu)\biggr].
\end{eqnarray}
Eq.\,$(\ref{total-probability-energy2})$ is independent from the position
${\vec X}_{\pi}$ of the initial state and depends upon the momenta and
the time difference $\text{T}=T_{\nu}-T_{\pi}$. Consequently although quantum
mechanics tells that the probability 
Eq.$(\ref{total-probability-energy2})$ should be applied to events of
one initial state of the pion, it is possible to  average over
different positions. The result is  obviously the same Eq.\,$(\ref{total-probability-energy2}  )$.

At $\text{T} \rightarrow \infty$, $\tilde{g}(\text{T},\omega_\nu)$ vanishes
and the decay rate per time of the energy $E_{\pi}$ is 
given  from the
first term of  Eq.\,$(\ref{total-probability-energy2})$ as
 \begin{eqnarray}
& &P/\text{T}=g^2 m_{\mu}^2
 D_0   {1 \over E_{\pi}}
 \frac{2\pi}{|\vec{P}_\pi|} {\pi}( m_\pi^2
 - m_\mu^2) \int_{E_{\nu,min}}^{E_{\nu,max}} dE_{\nu}\nonumber\\
& &=g^2 m_{\mu}^2 {1 \over 4\pi}\frac{m_{\pi}^2}{E_{\pi}}\left(1-{m_{\mu}^2
 \over m_{\pi}^2}\right)^2.
\end{eqnarray} 
The stationary value is independent from the wave packet size and is
consistent with \cite{Stodolsky}. This value furthermore agrees to the standard 
value obtained using the plane waves. Hence the large time limit of our 
result is equivalent to the known result. 

At  finite $\text{T}$, the diffraction term does not vanish and
Eq.\,$(\ref{total-probability-energy2})$ deviates from  the standard 
value obtained using the plane waves.      
The relative magnitude of the diffraction  term $\tilde
g(\text{T},\omega_{\nu})$ to  the normal 
term $G_0$ is  independent   from detection process. So we compute  
$\tilde g(\text{T},\omega_{\nu})$ and $G_0$ of 
Eq.~($\ref{probability-3}$) at the forward direction $\theta=0$ and the 
energy dependent total probability that is integrated over the neutrino
angle in the following.  
 \begin{figure}[t]%
\begin{center}
  \includegraphics[angle=-90,scale=.5]{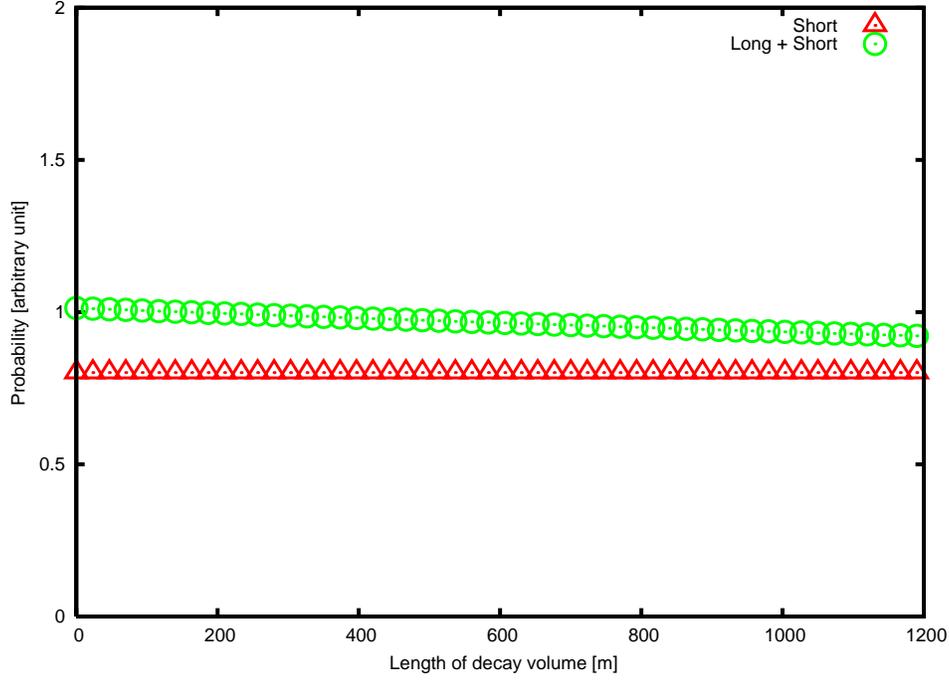}
   \end{center}
  \caption{The  total   probability integrated in the  neutrino angle per time at a finite distance L is 
given. The constant shows the short-range normal term and   the long-range term is written on top of the normal term.  The horizontal axis 
shows the distance in~[m] and the probability  is
  of arbitrary unit. Clear uniform excess is seen in the
 distance below 1200\,[m]. The neutrino mass, pion energy, neutrino energy are
  0.6~[eV/$c^2$], 4~[GeV], and 800~[MeV].}
 \label{fig:total-int-.6}
\end{figure}%
The function $\tilde g(\text{T},\omega)$ and $G_0$ are plotted in 
Fig.\,\ref{fig:virtual-pi-singular} for the mass of neutrino,
 $m_{\nu}=1\,[\text{eV}/c^2]$,  the pion energy $E_\pi = 4$\,[GeV], and the neutrino
 energy $E_\nu = 800$\,[MeV]. For the wave packet size of the neutrino, the size
 of the nucleus of the mass number $A$, $\sigma_{\nu}=
 A^{\frac{2}{3}}/m_{\pi}^2$ 
is used. The value becomes  $\sigma_{\nu}= 6.4/m_{\pi}^2$ for the ${}^{16}$O nucleus and this is used for the following
 evaluations.   From 
this figure it is seen   that there
 is an excess of the flux at short distance region $\text{L}<600$ [m] and the
 maximal excess is about $0.4$ at $\text{L}=0$. The slope at the origin $\text{L}=0$ is
 determined by $\omega_{\nu}$.
The slowly decreasing   term that  is generated from the singularity at
 the light cone has a finite magnitude. 

\begin{figure}[t]
   \begin{center}
   \includegraphics[angle=-90,scale=.5]{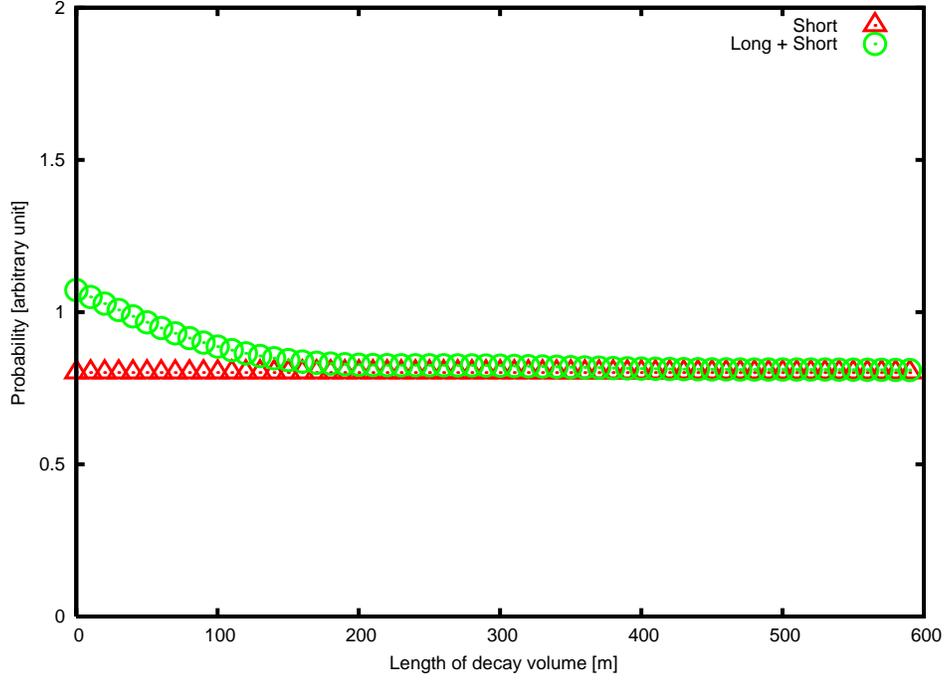}
   \end{center}
\caption{The  total  probability integrated in the angle per time at a 
finite distance L is 
given. The constant shows the short-range normal term and   the 
long-range term is written on top of the normal term.  The horizontal axis 
shows the distance in~[m] and the probability  is
  of arbitrary unit. Clear excess and decreasing behavior are  seen in the
 distance below 600~[m]. The neutrino mass, pion energy, neutrino energy are
  1~[eV/$c^2$], 4~[GeV], and 100~[MeV].}
 \label{fig:total-int-1-100}
\end{figure}

The total probability that is integrated over the neutrino angle
 Eq.\,$(\ref{total-probability-energy})$ is
 presented next. The probability for the  neutrino mass
 $m_{\nu}=1.0\,[\text{eV}/c^2]$ and the pion energy  $4$\,[GeV] and
 $4.5$\,[GeV] are given in  Fig.~$\ref{fig:total-int-1}$, and for the 
smaller neutrino mass 
$m_{\nu}=0.6\,[\text{eV}/c^2]$ is given in  Fig.\,$\ref{fig:total-int-.6}$. $G_0$ 
is unchanged with the distance but the long-distance term, $\tilde
 g(\text{T},\omega_{\nu})$, decreases  slowly with the distance than that  of
 $m_{\nu}=1\,[\text{eV}/c^2]$.   Hence the longer distance is necessary
 if the mass of the neutrino is even smaller. For the muon neutrino, it
 is impossible to measure the event at a energy lower than few $100$\,[MeV]. 
The electron neutrino is used then. Considering the situation for the
 electron neutrino, we present the total probability for the lower
 energies. The probability for the
 neutrino mass $m_{\nu}=1.0\,[\text{eV}/c^2]$ with the energy  $100$\,[MeV]
is given in  Fig.\,$\ref{fig:total-int-1-100}$.
The slowly decreasing component of the probability becomes more
 prominent with lower values.  Hence to observe this component, the 
experiment of the lower neutrino energy is more convenient.   
 \begin{figure}[t]
  \begin{center}
   \includegraphics[angle=-90,scale=.5]{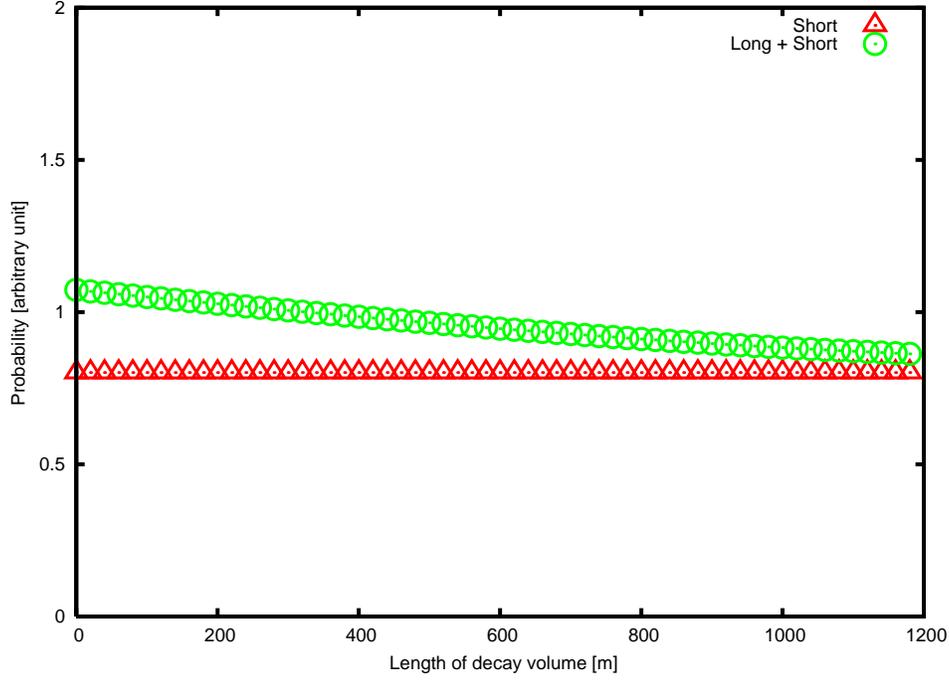}
  \end{center}
  \caption{The  neutrino  probability integrated in the neutrino angle 
per time at a finite distance L is 
given. The constant shows the short-range normal term and   the 
long-range term is written on top of the normal term.  The horizontal axis 
shows the distance in~[m] and the probability  is
  of arbitrary unit. Clear excess is seen in the
 distance below 1200~[m]. The neutrino mass, pion energy, neutrino energy are
  0.1~[eV/$c^2$], 4~[GeV], and 10~[MeV].}
 \label{fig:total-int-.1-10}
\end{figure}%

The typical length $\text{L}_0$ of the diffraction  term is  
\begin{eqnarray}
\text{L}_0~[\text{m}] ={2E_{\nu} \hbar c \over m_{\nu}^2 }= 400{E_{\nu}[\text{GeV}] \over
 m_{\nu}^2[\text{eV}^2/c^4]}.
\end{eqnarray}
By the observation of this component together with the neutrino's energy, the 
determination of the neutrino mass  may becomes possible. The neutrino's 
energy is measured with uncertainty $\Delta E_{\nu}$, which is of the 
order $0.1 \times E_{\nu}$. This uncertainty is $100$\,[MeV] for the energy
$1$\,[GeV] and is accidentally same order as that of the minimum uncertainty 
$\hbar/\delta x$ derived from  Eq.\,$(\ref{mfp-neutrino-N})$. The total probability for a larger
value of energy uncertainty is easily computed using
Eq.\,(\ref{probability-3}).   Figs.\,(3)-(7) show the length
dependence of the probability. If the mass is around $1\,[\text{eV}/c^2]$ the
excess of the neutrino flux of
about $20$ percent at the distance less than a few hundred meters is
found. In the  long-baseline neutrino oscillation experiments, the
neutrino flux at the near detectors has observed excesses of about $10-20$
percent
\cite{excess-near-detectorK2K,excess-near-detectorMini,excess-near-detectorMino}. We 
believe this is connected with the excesses found in this
paper. We use mainly $m_{\nu}=1\,[\text{eV}/c^2]$ throughout this paper.
 \begin{figure}[t]
  \begin{center}
   \includegraphics[angle=-90,scale=.5]{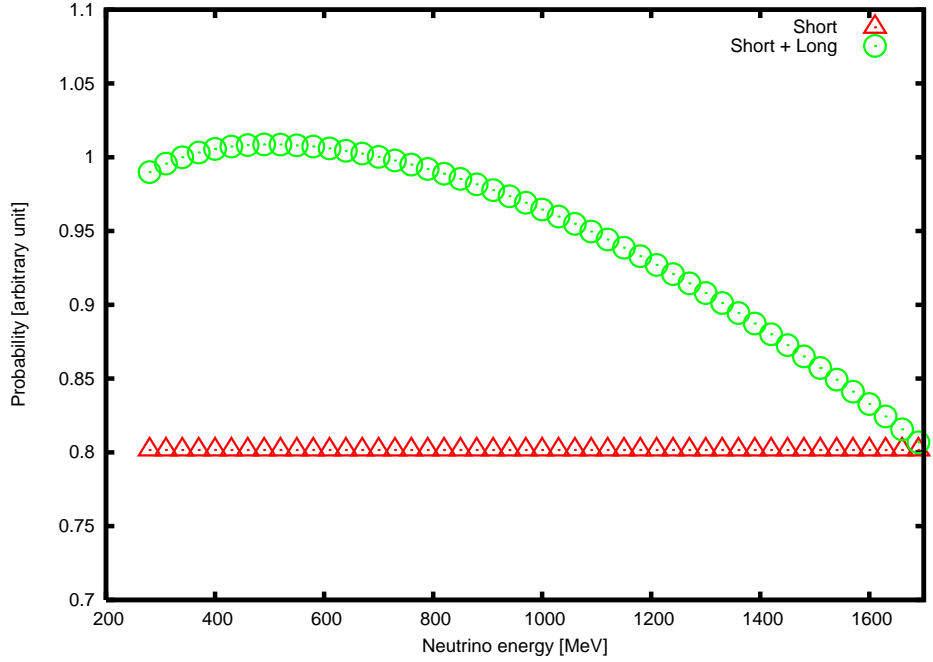}
  \end{center}
  \caption{The energy dependence of the probability integrated 
  in the angle at distance $\text{L} = 100$ [m] is 
given. The lower curve  shows the short-range normal term and   the long-range term is added  on top of the normal term.  The horizontal axis 
shows the neutrino energy in~[MeV] and the probability  is
  of arbitrary unit.  The neutrino mass and pion energy are
  1.0~[eV/$c^2$] and 4~[GeV].}
 \label{fig:total-ene-1}
\end{figure}%
Because the probability has a constant  term and the T-dependent 
term, the T-dependent   term is extracted  easily 
by subtracting the constant term from the total probability. The slowly 
decreasing  component decreases with the scale determined by the
neutrino's mass and the energy.  
Although the excess of the flux would be found, the decreasing behavior  
becomes difficult to observe if the mass is less than $0.1\,[\text{eV}/c^2]$ for 
the muon neutrino. In this case,  the 
electron neutrino is useful. The electron neutrino is produced in the
decays of the muon, neutron, K-meson, and nucleus. In these decays the 
present mechanism works. So we plot the figure for $m_{\nu}=0.1
\,[\text{eV}/c^2]$, $E_{\nu}=10\,[\text{MeV}]$ in
Fig.\,$\ref{fig:total-int-.1-10}$.
A decreasing part is clearly seen. So
in order to  observe    the slow decreasing behavior for the small
neutrino  mass less than or about the same as $0.1\,[\text{eV}/c^2]$, the electron
neutrino should be used. The decay of the muon and others will be
studied in a forthcoming paper.
In Fig.\,$\ref{fig:total-ene-1}$ the energy dependence of the total
probability is given. The  long-range term has a different form and has
a maximum at the neutrino energy $E_{\nu}\approx {1/3 E_{\nu,max}}$.
\subsubsection{Wide distribution of pion momentum  }
When the momentum distribution $\rho_{exp}({\vec p}_{\pi})$ of the
initial pion is known,  the energy-dependent probability of the neutrino
is given by  Eq.\,$(\ref{total-probability-energy})$.
 Eq.\,$(\ref{total-probability-energy})$ is also independent from the position
${\vec X}_{\pi}$ and depends upon the momentum of the pion and of the
neutrino  and the time difference $\text{T}=T_{\nu}-T_{\pi}$. In the experiments, the
pion's position is not measured and the average over the position is made. 
This  average probability  agrees to 
Eq.$(\ref{total-probability-energy})$. The probability varies slowly 
with  the pion's momentum and is regarded constant in the energy range 
of order $100$\,[MeV]. So the experimental observation of the diffraction  
term is quite easy.
   
\subsection{On the universality of the long-distance correction}
 The long-distance correction of the neutrino detection
probability  was obtained using the wave 
packets. The  correction is decreasing 
extremely slowly with the distance. 
The slope is  determined  only from the mass and energy of 
the neutrino and is  independent from details of other parameters of 
the system such as the size or shape of the wave packets and the 
position of  the final wave packets and others. Hence this  is 
the genuine property of the wave function 
$|\text {muon,~neutrino}(t)\rangle$, of Eq.\,$(\ref{state-vector} )$. The 
slope of this term  has the universal property and is capable of measuring in 
experiments.
\subsubsection{Muon in pion decay}
When the muon is observed in the same processes, the anomalous behavior 
is determined by the mass and energy of the muon as
${m_{\mu}^2/2E_{\mu}}$. 
Since  the muon mass is  larger than the neutrino mass
by $10^8$,  the typical length for the muon is smaller than that of 
the neutrino by $10^{-16}$. For
the muon of energy 1\,$[\text{GeV}]$, the length is order $10^{-12}$\,[m].
This value is a microscopic size. So the muon probability becomes
constant in this distance and has no observable long-range effect. The
muon from the pion decay has no coherence effect and can be treated
incoherently.
\section{Summary and implications}

 In this paper, we  studied the interference of the neutrino in high
 energy pion decay. The position-dependent neutrino flux  was computed and  
the  new long-distance component of  universal property that  is
 insensitive to  the pion's initial conditions was found to
 exist in addition  to the normal in-coherent term. The pion's coherence
 length estimated in Section 2 shows that this is  long  enough for
 the new universal  term  to be observed.
Since the new term 
reflects the interference of the neutrino,  it decreases with the
 distance in the universal
 manner that depends upon the absolute value of the neutrino mass.
 
The present interference phenomenon is caused by the light-cone 
singularity of the correlation function of the pion
 and muon system  $\Delta_{\pi,\mu}(\delta t,\delta\vec{x})$ and the 
slow phase
 $\bar{\phi}_c$ of the neutrino wave function,
Eq.\,$(\ref{light-phase})$ 
These 
two ingredients are characteristic phenomena of the relativistic 
invariant system. The
 light-cone singularity is formed from the waves of the infinite
 momentum, which have always the light velocity. 
The slow phase of the neutrino wave packet is the outcome of the
 cancellation between the time and space oscillations. The phase is 
determined by the difference of space-time coordinates $(\delta t,\delta
 {\vec x})$ and the
 central values of the energy and momentum,  as
${E({\vec p}\,)\delta t-{\vec p}\!\cdot\! \delta {\vec x}}$, where the energy $E({\vec p}\,)$ is
 given by $\sqrt {{\vec p}^{\, 2}+m^2}$. 
When the  difference of positions $\delta {\vec x}$ is 
moving with the light velocity in the parallel direction to the momentum ${\vec
p}$, ${\vec v}_c=c({\vec p}/p)$, as $\delta {\vec x}={\vec v}_c \delta t$, then     
the total  phase becomes $E_{\nu}\delta t-{\vec p}_\nu\!\cdot\! \delta {\vec
 x}=(m_\nu^2/2E_{\nu}) \delta t$.
 The angular velocity becomes the small value $m_\nu^2/2E_{\nu}$ due to the relativistic invariance and makes the interference
 phenomenon long-range. 
The new term in the probability 
decreases   slowly with the distance in the universal manner determined
by the mass and energy of the neutrino as $m_{\nu}^2/2E_{\nu}$. 
This  form is independent from the details of wave packet shape and 
parameters as far as the reality of the neutrino wave function $\tilde
w({\vec x})$ is satisfied, which is ensured from the invariance under
the time inversion.  The relative magnitude of this component is not
universal and depends upon  the size of wave packet. 
Based on the estimation   of the size, we found that the magnitude of 
the new universal term is sizable for the measurement.
Since the slope of  the probability, Eq.\,$(\ref{probability-3})$, is 
determined by 
the mass and energy of the neutrino, the 
absolute value of the neutrino mass would be found from the neutrino 
interference experiments.

The new component shall be observed as the
excesses of the neutrino flux in the ground experiments. The excesses of 
the neutrino flux at the macroscopic short distance region of the order 
of a few hundred meters were 
computed and shown in
Figs. (3)-(8). From these figures, the excesses are  not large but
are sizable magnitudes. Hence  these excesses  shall be observed in these  
distances. 
Actually fluxes measured in the near detectors of the long-baseline 
experiments of K2K \cite{excess-near-detectorK2K} and MiniBooNE
\cite{excess-near-detectorMini}  may show  excesses of about
$10-20$ percent of the Monte Carlo estimations. Monte Carlo estimations of
the fluxes are obtained using naive decay probabilities and do not have
the coherence effects we presented in the present work. So the excess of
these experiments may be related with the excesses due to
interferences. The excess is not clear in MINOS 
\cite{excess-near-detectorMino}. With more
statistics, qualitative analysis  might become  possible to test the new 
universal term on the neutrino  flux at the finite distance.   From
Figs. (3)-(8),
 if the mass is in the range from $0.1\,[\text{eV}/c^2]$ to $2\,[\text{eV}/c^2]$, the 
near detectors  at T2K, MiniBooNE, MINOS and other experiments  might 
be able to measure these signatures. The absolute value of the mass 
could be found then.    
It is worthwhile at the end to summarize the reasons why the 
interference term of the long-distance behavior emerges in the pion
decay and is computed with the wave packet representation. The
connection of the long-distance interference phenomenon of the neutrino with 
the Heisenberg's uncertainty relation is also addressed here. 

  Due to  relativistic invariance, the correlation function
$\Delta_{\pi,\mu}(\delta t, \delta\vec{x})$
 has a singularity near  $\lambda=0$,
which is extended to  large distance $|\delta {\vec x|} \rightarrow
\infty$, as was shown in Eq.\,(\ref{singular-function-f}). This is one of the  features of 
relativistic quantum fields and is one reason why the long-range
correlation emerged.  
For a non-relativistic system, on the other hand,
the same calculation for stationary states is
made by,
\begin{eqnarray}
\int d {\vec k} \langle {\vec x}_1| {\vec k} \rangle \langle {\vec k}|{\vec
x}_2 \rangle = \delta({\vec x}_1-{\vec x}_2),
\end{eqnarray}
and the only one point $\delta {\vec x}=0$ satisfies the condition.
 Long-range correlation is not generated. The  rotational invariant  three-dimensional space 
is compact but the Lorentz invariant  four-dimensional space is non-compact.   
So it is quite natural for the non-relativistic system not to have the 
long-range correlation that the relativistic system has. The light-cone 
singularity is the peculiar  property of the relativistic system. 

Heisenberg uncertainty relation is slightly modified in the wave along
the light cone. The 
neutrino wave function behaves at  large-distance along the
light-cone region 
in the form
\begin{eqnarray}
& &\psi_{\nu}(t,{\vec x})=f{e^{i(E_{\nu}t-{\vec p}_{\nu}\cdot{\vec x})} \over x}= f{e^{i{m_{\nu}^2 \over
2E_{\nu}}t}\over ct},
\end{eqnarray}
where $f$ has no dependence on the distance $|\vec{x}|$. Consequently the
uncertainty relation between the energy width $\delta E$ and the time
interval $\delta t$ becomes
\begin{eqnarray}
\delta t \delta{m_{\nu}^2 \over 2E_{\nu}} =\delta t \delta E 
\times{1 \over 2} \left({m_{\nu} \over E_{\nu}}\right)^2 \approx \hbar.
\end{eqnarray}
The ratio $({m_{\nu}/E_{\nu}})^2$ is of the order $10^{-18}$ 
and $\delta t$ becomes macroscopic  even if the energy width
$\delta E$ is  microscopic of order 100 [MeV]. For instance 
if the pion Compton wave length, $\lambda_{\pi}$, is used for the 
microscopic length, then $c \delta
t$ becomes 
\begin{eqnarray}
10^{18}\times \lambda_{\pi} \approx 10^3 m,
\end{eqnarray} 
 which is about the distance between the pion source and the near 
detector in fact. So  interference effect of the present paper
appears in this distance and is observable using the apparatus of much
smaller size.     

The position-dependent probability was  computed easily with the use
of the wave packet representation. Any representation   
is equivalent to plane waves as far as the
whole complete set of functions  is used.  
The ordinary probability of the transition process is defined from the 
states at  $t=\pm \infty$. The  normal scattering
amplitude is the overlap between  the in-state at $t=-\infty$ and 
out-state at $t=\infty$, and the space-time
coordinates are 
integrated from $-\infty$ to $\infty$ and the energy and momentum of the
final state is the same as that of the initial state.
Hence in the pion decay, the momentum  of the muon in the final state 
of the ordinary scattering  experiments are bounded.  
So the infinite momentum is not included in the muon of the 
final state.  From  these amplitudes, the  amplitudes and probability 
at the finite-time interval are  neither computable nor obtained. 
In the wave packet formalism, on the  other hand, it is possible to 
compute the amplitude and probability at  finite-time interval directly. 
Energy and momentum conservation  is slightly violated in this 
amplitude and the states of  infinite
momentum  couple and give the light-cone singularity to the correlation
function $\Delta_{\pi,\mu}({\delta x})$. The contribution from these states
vanishes at  infinite time and  these states  do not contribute to the 
probability measured at infinite distance, i.e., ordinary S-matrix.
Thus the light-cone singularity gave  the interference term, which  
ended up as  the finite-time interval effect of the neutrino probability.
 Wave packet formalism gives  new information that  is  not
 calculable 
in the standard scattering
amplitude. Hence our calculation does not 
contradict with the ordinary calculation of the S-matrix in momentum 
representation but has  advantage of giving a new universal physical 
quantity directly.

Since characteristic small phase of the relativistic wave along the light
cone and the light-cone singularity are derived from  Lorentz
invariance, there would be similar  effect if there exists other  light
particle. Axion is a  possible candidate of light particle and 
might show the long-distance interference phenomenon.

In this paper we ignored  higher order effects such as pion life 
time and pion mean free path in studying the quantum effects. We will
study these problems  and other large scale physical phenomena 
of low energy neutrinos in subsequent papers.

    
\section*{Acknowledgements}
One of the authors (K.I) thanks Dr. Nishikawa for useful discussions on 
the near detector of T2K experiment, Dr. Asai, Dr. Mori, and Dr. Yamada
for useful discussions on interferences. This work was partially supported
by a Grant-in-Aid for Scientific Research(Grant No. 19540253 ) provided
by the Ministry of Education,
Science, Sports and Culture,and a Grant-in-Aid for Scientific Research on 
Priority Area ( Progress in Elementary Particle Physics of the 21st
Century through Discoveries  of Higgs Boson and Supersymmetry, Grant 
No. 16081201) provided by 
the Ministry of Education, Science, Sports and Culture, Japan. 
\\
{}

\appendix
\def\thesection{Appendix \Alph{section}}
\def\thesubsection{\Alph{section}-\Roman{subsection}}

\section{Light-cone singularity }\label{App:OPE}

\subsection{Light-cone singularity and small neutrino mass }
A conjugate momentum to $(\delta t, \delta {\vec x})$ is $p_{\pi}
-p_{\mu}$ from Eq.~$(\ref{pi-mucorrelation})$ and 
the invariant square of this momentum becomes 
\begin{align}
(p_{\pi}-p_{\mu})^2=m_{\pi}^2+m_{\mu}^2-2\left(\sqrt{\vec{p}_{\pi}^{\,2}+m_{\pi}^2}
 \right)\left(\sqrt{\vec{p}_{\mu}^{\,2} + m_{\mu}^2} \right)+2|\vec{p}_{\pi}||\vec{p}_{\mu}|\cos \theta,
\end{align} 
where  $\theta$ is the angle between the pion and muon
momenta. The invariant  vanishes when the cosine of angle 
becomes    
\begin{align}
\cos \theta_c&={-m_{\pi}^2-m_{\mu}^2+2\left(\sqrt{\vec{p}_{\pi}^{\,2}+m_{\pi}^2}
 \right)\left(\sqrt{\vec{p}_{\mu}^{\,2}+m_{\mu}^2} \right) \over 2|\vec{p}_{\pi}||\vec{p}_{\mu}|} \nonumber\\
 &=1-{{m_{\pi}}^2 \over 2|\vec{p}_{\pi}||\vec{p}_{\mu}|}\left(1-{|\vec{p}_{\mu}| \over
  |\vec{p}_{\pi}|}\right)-{{m_{\mu}}^2 \over 2|\vec{p}_{\pi}||\vec{p}_{\mu}|}\left(1-{|\vec{p}_{\pi}| \over
  |\vec{p}_{\mu}|}\right)+\text{ small~ terms}.
\end{align}
This equation has a solution in a finite muon momentum region for a
given pion momentum. 

\subsection{ Long-range correlation for general wave packets  }
\subsubsection{ Non-Gaussian wave packet}
Non-Gaussian wave packets were studied in the general manner in the 
text and the universal behavior of the phase was obtained. In this
appendix the explicit forms of the wave packets are studied. It is reconfirmed
that the long-range component of the probability  at around  $t={2 \pi
E_\nu/m_\nu^2}$ becomes the universal form.

{\bf type 1}

One way to express the non-Gaussian wave packet is to multiply 
 Hermitian polynomials and to write the amplitude in the form 
\begin{eqnarray}
  \frac{N_{\nu}}{(2\pi)^{\frac{3}{2}}}\int d{\vec
 k}_{\nu}e^{-{\sigma_{\nu} \over 2}({\vec k}_{\nu}-{\vec
 p}_{\nu})^2}H_n(\sqrt{\sigma_{\nu}}({\vec k}_{\nu}-{\vec p}_{\nu}))
e^{i\left(E({\vec k}_{\nu})(t-T_{\nu})-{\vec
 k}_{\nu}\cdot({\vec x}-{\vec X}_{\nu})\right)}.  
\end{eqnarray}
where $H_n$ is assumed to be real  in order the wave packets 
to preserve the time reversal symmetry  and an even function of ${\vec
k_{\nu}}-{\vec p}_{\nu}$ in order the wave packets 
to preserve parity , Eqs.\, $(\ref{time-reversal})$ and $(\ref{parity})$.

Since we study  symmetric wave packets, it is sufficient to prove the
simplest case
\begin{eqnarray}
H_n= {\sigma_{\nu}}({\vec k}_{\nu}-{\vec p}_{\nu})^2. 
\label{quadratic-form}
\end{eqnarray}
The spreading effect was studied in the previous appendix and does not
change the final result. So we ignore the spreading effect here. 
The momentum integration Eq.\,$(\ref{amplitude})$ is replaced with 
\begin{eqnarray}
\int d{\vec k}_\nu e^{-{\sigma_{\nu} \over 2}({\vec k}_\nu-{\vec p}_\nu)^2}\sigma_{\nu}({\vec
 k}_\nu-{\vec p}_{\nu})^2e^{i(E({\vec k}_\nu)(t-T_{\nu})-{\vec k}_\nu\cdot({\vec x}-{\vec
 X}_{\nu}))}. 
\end{eqnarray}
After the straightforward calculations we have this integral in the form,
\begin{eqnarray}
& &\sigma_{\nu}e^{i(E({\vec p}_\nu)(t-T_{\nu})-{\vec p}_\nu\cdot({\vec x}-{\vec X}_{\nu})) -{1 \over
 2\sigma_{\nu}}({\vec x}-{\vec X}_{\nu}-{\vec v}_\nu(t-T_{\nu}))^2} \int
 d{\vec k}_\nu e^{-{\sigma_{\nu} \over 2}\left({\vec k}_\nu -{\vec p}_\nu+{i \over
 \sigma_{\nu}}\{{\vec x}-{\vec X}_{\nu}-{\vec v}_\nu(t-T_{\nu})\}\right)^2} \nonumber \\
& &\times \Biggr[\left({\vec k}_\nu-{\vec p}_\nu+{i \over
 \sigma_{\nu}}\{{\vec x}-{\vec X}_{\nu}-{\vec v}_\nu(t-T_{\nu})\}\right)^2-{1 \over \sigma_{\nu}^2}\{{\vec
 x}-{\vec X}_{\nu}-{\vec v}_\nu(t-T_{\nu})\}^2 \nonumber\\
& &+{2i \over \sigma_{\nu}}\left({\vec k}_\nu-{\vec p}_\nu+{i \over
 \sigma_{\nu}}\{{\vec x}-{\vec X}_{\nu}-{\vec v}_\nu(t-T_{\nu})\}\right)\{{\vec
 x}-{\vec X}_{\nu}-{\vec v}_\nu(t-T_{\nu}) \} \Biggr]  \nonumber\\
& &=\left({2\pi \over \sigma_{\nu}}\right)^{\frac{3}{2}} e^{i(E({\vec p}_\nu)(t-T_{\nu})-{\vec p}_\nu\cdot({\vec x}-{\vec
 X}_{\nu})) -{1 \over 2\sigma_{\nu}}({\vec x}-{\vec X}_{\nu}-{\vec v}_\nu(t-T_{\nu}))^2}\nonumber \\
& &~~~\times \left(3-{1 \over \sigma_{\nu}}\{{\vec x}-{\vec X}_{\nu}-{\vec
		v}_\nu(t-T_{\nu})\}^2 \right).
\end{eqnarray}

Next the integral Eq.\,({\ref{singular-correlation}}) is studied. This
becomes for the non-Gaussian wave packet to the  integral   
\begin{align}
&\tilde J_{\delta({\lambda})}=N_\nu^2 \int d{\vec x}_1 d {\vec x}_2 e^{i\phi(\delta x)  -{1 \over
 2\sigma_{\nu}}({\vec x}_1-{\vec X}_{\nu}-{\vec v}_\nu(t_1-T_{\nu}))^2 -{1 \over
 2\sigma_{\nu}}({\vec x}_2-{\vec X}_{\nu}-{\vec v}_\nu(t_2-T_{\nu}))^2}{1 \over 4\pi} \delta(\lambda)\nonumber\\
&\times \left(3-{1 \over \sigma_{\nu}}\{{\vec x}_1-{\vec X}_{\nu}-{\vec
		v}_\nu(t_1-T_{\nu})\}^2 \right)
\left(3-{1 \over \sigma_{\nu}}\{{\vec x}_2-{\vec X}_{\nu}-{\vec
		v}_\nu(t_2-T_{\nu})\}^2 \right),
\end{align}
and is written 
  by using the center coordinate 
 ${X}^\mu={{x}_1^\mu + {x}_2^\mu \over 2}$ and relative 
coordinate ${r}^\mu={x}_1^\mu - {x}_2^\mu$ in the form
\begin{align}
&N^2 \int d{ \vec X} d { \vec r} e^{i\phi(\delta x)  -{1 \over
 \sigma_\nu} \tilde {\vec X}^2-{1 \over {4 \sigma_\nu}}\tilde{\vec{r}}^{\,2}
 }{\delta(\lambda) \over 4\pi}   \nonumber \\ 
&\ \ \ \times
\Biggl[9-{3 \over \sigma}\left(2\tilde
 {\vec X}^2  +\frac{1}{2}\tilde {\vec r}^{\,2}\right)
+\frac{1}{\sigma_\nu^2}\left(\tilde{\vec X}^4 -\frac{1}{2}\tilde{\vec X}^2\tilde{\vec r}^{\,2}+\frac{1}{16}\tilde {\vec r}^{\,4}\right)\Biggr], 
\end{align}
where
\begin{eqnarray}
\tilde {\vec X}={\vec X}-{\vec X}_\nu-{\vec v}_\nu(X^0-T_\nu),\ \tilde {\vec r}={\vec r}-{\vec v}r^0\nonumber.
\end{eqnarray}
The integration on ${\vec X}$ and ${\vec r}$ are made and we have the 
final result  
\begin{align}
\tilde J_{\delta({\lambda})}&=
N_\nu^2(\sigma_\nu\pi)^{\frac{3}{2}}\sigma_\nu {1 \over 2r^0}
 e^{i(E-p_\nu)r^0}\nonumber\\
& \ \ \times \left[-{13 \over
 4}+{9 \over 4}{1
 \over \sigma_\nu}(1-v_\nu)^2(r^0)^2 +O((1-v_\nu)^4(r^0)^4 )\right].
\end{align}

Thus the phase factor has the same universal form as the Gaussian wave 
packet and the correction is determined by the small 
parameter $(1-v_\nu)^2(r^0)^2 $ in the form
\begin{eqnarray}
{1 \over \sigma_\nu}(1-v_\nu)^2(r^0)^2=\left({1 \over E_{\nu} \sigma_\nu}\right)^2\left({m_{\nu}^2
 \over 2E_{\nu}}r^0\right)^2,
\label{non-gaussian-correction}
\end{eqnarray}
hence the correction is negligible at high energy. 

We have proved that  the correlation function of the non-Gaussian wave 
packet has the same slow phase and long-range term as the Gaussian 
wave packet and the small correction becomes negligible for the simplest case
Eq.\,$\ref{quadratic-form}$.  Hence for any polynomials $H_n$ that are
invariant under the time and space inversions,  the correlation function
has the same long-range term and small negligible corrections. 

{\bf type 2}

Another way to write the non-Gaussian wave packet is to use a function 
$\alpha({\vec p}\,)$, and to write  
\begin{eqnarray}
\label{non-gaussian-alpha}
  \frac{N_{\nu}}{(2\pi)^{\frac{3}{2}}}\int d{\vec
 k}_{\nu}e^{-\alpha({\vec k}_{\nu})+i\left(E({\vec k}_{\nu})(t-T_{\nu})-{\vec
 k}_{\nu}\cdot({\vec x}-{\vec X}_{\nu})\right)}.
\end{eqnarray}
The large $t=T$ behavior is found by the stationary momentum which 
satisfies the equation 
\begin{eqnarray}
{\partial \over \partial k_{\nu,i}}\alpha|_{{\vec k}_\nu={\vec p}_\nu}=0.
\end{eqnarray}
Symmetric real wave packet is assumed also here from parity and time
reversal invariances of the wave packets and we write,
\begin{eqnarray}
\alpha({\vec k}_\nu)=\alpha({\vec p}_\nu)+{({\vec k}_\nu-{\vec p}_\nu\,)^2 \over 2}
 \sigma_\nu +(k_\nu-p_\nu)_i^2(k_\nu-p_\nu)_j^2 C{ij}+\cdots,
\end{eqnarray}
where the $\sigma$ and $C_{ij}$ are real numbers. 
The momentum integration of Eq.\,$(\ref{non-gaussian-alpha})$ becomes the form
\begin{eqnarray}
& &  \frac{N'_{\nu}}{(2\pi)^{\frac{3}{2}}}\int d{\vec
 k}_{\nu}e^{-{\sigma_\nu \over 2}({\vec k}_{\nu}-{\vec p}_{\nu})^2+i\left(E({\vec k}_{\nu})(t-T_{\nu})-{\vec
 k}_{\nu}\cdot({\vec x}-{\vec X}_{\nu})\right)} e^{-( ((k_\nu-p_\nu)_i)^2((k_\nu-p_\nu)_j)^2 C{ij})}\nonumber  \\
& &=\frac{N'_{\nu}}{(2\pi)^{\frac{3}{2}}}\int d{\vec
 k}_{\nu}e^{-{\sigma_\nu \over 2}({\vec k}_{\nu}-{\vec p}_{\nu})^2+i\left(E({\vec k}_{\nu})(t-T_{\nu})-{\vec
 k}_{\nu}\cdot({\vec x}-{\vec X}_{\nu})\right)}  \nonumber\\
& &\times(1- ((k_\nu-p_\nu)_i)^2((k_\nu-p_\nu)_j)^2 C{ij}) . 
\end{eqnarray}
The correction to the Gaussian wave packet is generated by the higher
order terms of  ${\vec k}_\nu-{\vec p}_\nu$ in the right hand side and is
treated in a same way as the previous type 1 case. Hence this integral 
has the  leading long-range term which is equivalent to that of 
the Gaussian wave packet and the negligible correction 
expressed by Eq.\,$(\ref{non-gaussian-correction})$.

For studying the asymptotic behavior at $t-T_\nu \rightarrow \infty $ we solve
the stationarity equation,  
\begin{eqnarray}
{\partial \over \partial k_{\nu,i}}\left[\alpha({\vec k}_\nu)-i\{E({\vec k}_\nu)(t-T_\nu)-{\vec
 k}_\nu\cdot({\vec x}-{\vec X}_\nu)\}\right]=0
\end{eqnarray}
and expand the integral around the stationary momentum. The wave 
in the transverse direction to this  momentum spreads but spreading is
very small in the longitudinal direction \cite{Ishikawa-Shimomura}. From the result of the 
previous appendix, the final result is the same and so  is not presented here.

$\bf type~ 3$

In the type 1 and 2 the time reversal and parity symmetries  are assumed
for the wave packet shape. If these symmetries are not required, the
function $H_n$ or $\alpha$ has an imaginary part. In this case, the
correlation function has a correction term in the order $(1-v)(t_1-t_2)$
and  this term is expressed 
\begin{eqnarray}
(1-v_\nu)(t_1-t_2)={1 \over E_\nu}{m_{\nu}^2 \over 2E_{\nu}}(t_1-t_2)
\end{eqnarray} 
hence the correction term vanishes at the high energy. With a suitable
parameter,  the universal form of the slowly decreasing component of the
probability of the present work may become observable even in arbitrary
system. The Lorentz invariant form of the energy dependent phase of the 
wave packet and the light-cone singularity of the pion and muon decay
vertex give this universal behavior. 


\begin{thebibliography}{}
\bibitem{J-J} The last form of the interaction Hamiltonian is used in
	the text. If the first expression  
of using the (V-A) current and the derivative of the pion field is used, the
	magnitude of the long-distance term is slightly modified.
\end{thebibliography}
\end{document}